\documentclass{ieeeaccess}
\usepackage{cite}
\usepackage{amsmath,amssymb,amsfonts}
\usepackage{algorithmic}
\usepackage{graphicx}
\usepackage{textcomp}

\usepackage{bm}
\makeatletter
\AtBeginDocument{\DeclareMathVersion{bold}
\SetSymbolFont{operators}{bold}{T1}{times}{b}{n}
\SetSymbolFont{NewLetters}{bold}{T1}{times}{b}{it}
\SetMathAlphabet{\mathrm}{bold}{T1}{times}{b}{n}
\SetMathAlphabet{\mathit}{bold}{T1}{times}{b}{it}
\SetMathAlphabet{\mathbf}{bold}{T1}{times}{b}{n}
\SetMathAlphabet{\mathtt}{bold}{OT1}{pcr}{b}{n}
\SetSymbolFont{symbols}{bold}{OMS}{cmsy}{b}{n}
\renewcommand\boldmath{\@nomath\boldmath\mathversion{bold}}}
\makeatother

\def\BibTeX{{\rm B\kern-.05em{\sc i\kern-.025em b}\kern-.08em
    T\kern-.1667em\lower.7ex\hbox{E}\kern-.125emX}}

\usepackage{booktabs}
\usepackage{hyperref}
\usepackage[mode=buildnew]{standalone}
\usepackage{cleveref} 
\usepackage{url}

\usepackage{xspace}
\usepackage[%
    font={small,sf},
    labelfont=bf,
    format=hang,    
    format=plain,
    margin=0pt,
    width=0.8\textwidth,
]{caption}
\usepackage[list=true]{subcaption}

\usepackage{pifont}
\newcommand{\cmark}{\ding{51}}%
\newcommand{\xmark}{\ding{55}}%

\newcommand{\xeon}{Intel\textregistered\xspace Xeon\textregistered\xspace}

\newcommand{\reint}{\mbox{\tt re\_32}\xspace}
\newcommand{\reiv}{\mbox{\tt re\_iv}\xspace}
\newcommand{\reans}{\mbox{\tt re\_ans}\xspace}

\newcommand{\euX}{{\tt eu-2005\/}\xspace}
\newcommand{\hollywoodX}{{\tt hollywood-2009\/}\xspace}
\newcommand{\inX}{{\tt in-2004\/}\xspace}  
\newcommand{\ljournalX}{{\tt ljournal-2008\/}\xspace}  
\newcommand{\indochinaX}{{\tt indochina-2004\/}\xspace}  

\newcommand{\ukX}{{\tt uk-2002\/}\xspace}  
\newcommand{\arabicX}{{\tt arabic-2005\/}\xspace}  
\newcommand{\ukXX}{{\tt uk-2005\/}\xspace}  
\newcommand{\itX}{{\tt it-2004\/}\xspace}  

\long\def\ignore#1{}

\newcommand{\perf}{{\tt perf\/}\xspace}

\newcommand{\sy}[2]{\langle{#1,\!#2}\rangle}
\newcommand{\sz}[1]{\$}  
\newcommand{\Val}{\mathop{\mathsf{eval}}\nolimits}
\newcommand{\Oh}{\mathcal O}

 \usepackage[usenames,dvipsnames]{xcolor}

\newcommand{\pferragina}[1]{\textcolor{red}{{\bf PF: }#1}}

\newcommand{\ftosoni}[1]{\textcolor{blue}{{\bf FT: }#1}}

\usepackage{tabularx}

\begin{document}
\history{Date of publication xxxx 00, 0000, date of current version xxxx 00, 0000.}
\doi{10.1109/ACCESS.2024.0429000}

\title{
Toward Greener Matrix Operations by Lossless Compressed Formats}
\author{
    \uppercase{Francesco Tosoni}\authorrefmark{1},
    \uppercase{Philip Bille}\authorrefmark{2},
    \uppercase{Valerio Brunacci}\authorrefmark{3}
    \uppercase{Alessio De Angelis}\authorrefmark{3},
    \uppercase{Paolo Ferragina}\authorrefmark{4},
    \uppercase{Giovanni Manzini}\authorrefmark{1},
}

\address[1]{Computer Science Department, University of Pisa, Polo Fibonacci, Building C, 3rd floor, L.go B. Pontecorvo 3, 56127 Pisa PI, Italy (e-mail: francesco.tosoni@di.unipi.it, giovanni.manzini@unipi.it)}
\address[2]{DTU Compute, Technical University of Denmark, DK-2800 Kgs. Lyngby, Denmark, (e-mail: phbi@dtu.dk)}
\address[3]{Engineering Department, University of Perugia, Via G. Duranti 93, 06125 Perugia PG, Italy (e-mail: alessio.deangelis@unipg.it, valerio.brunacci@dottorandi.unipg.it)}
\address[4]{EMbeDS Department, Sant'Anna School of Advanced Studies, Piazza Martiri della Libertà 33, 56127 Pisa PI, Italy (e-mail: paolo.ferragina@santannapisa.it)}

\tfootnote{This work was supported by the European Union -- Horizon 2020 Program under the scheme ``INFRAIA-01-2018-2019 -- Integrating Activities for Advanced Communities'', Grant Agreement n.~871042, ``SoBigData++: European Integrated Infrastructure for Social Mining and Big Data Analytics'', by the NextGenerationEU -- National Recovery and Resilience Plan (Piano Nazionale di Ripresa e Resilienza, PNRR) -- Project: ``SoBigData.it - Strengthening the Italian RI for Social Mining and Big Data Analytics'' -- Prot.~IR0000013 -- Avviso n.~3264 del 28/12/2021, by the spoke ``FutureHPC \& BigData'' of the ICSC -- Centro Nazionale di Ricerca in High-Performance Computing, Big Data and Quantum Computing funded by European Union -- NextGenerationEU -- PNRR, by the Independent Research Fund Denmark, DFF-9131-00069B).}

\corresp{Corresponding author: Francesco Tosoni (e-mail: francesco.tosoni@di.unipi.it).}

\begin{abstract}
Sparse matrix-vector multiplication (SpMV) is a fundamental operation in machine learning, scientific computing, and graph algorithms. In this paper, we investigate the space, time, and energy efficiency of SpMV using various compressed formats for large sparse matrices, focusing specifically on Boolean matrices and real-valued vectors.

Through extensive analysis and experiments conducted on server and edge devices, we found that different matrix compression formats offer distinct trade-offs among space usage, execution time, and energy consumption. Notably, by employing the appropriate compressed format, we can reduce energy consumption by an order of magnitude on both server and single-board computers. Furthermore, our experiments indicate that while data parallelism can enhance execution speed and energy efficiency, achieving simultaneous time and energy efficiency presents partially distinct challenges. Specifically, we show that for certain compression schemes, the optimal degree of parallelism for time does not align with that for energy, thereby challenging prevailing assumptions about a straightforward linear correlation between execution time and energy consumption.

Our results have significant implications for software engineers in all domains where SpMV operations are prevalent. They also suggest that similar studies exploring the trade-offs between time, space, and energy for other compressed data structures can substantially contribute to designing more energy-efficient software components.
\end{abstract}


\titlepgskip=-21pt

\maketitle

\section{Introduction}
\label{sec:introduction}
\PARstart{T}{he} emergence of Machine Learning (ML) and, in particular, Deep Learning algorithms has brought about significant advancements in computer vision (e.g., image classification~\cite{Kaiming:2016, Szegedy:2015}), natural language processing (e.g., text classification~\cite{GPT3-expensive, fasttext-models, bert-pre-training}), speech recognition, information retrieval, and more. These advancements have led to increased network complexity, more parameters, greater demands for training resources, and longer prediction latency. 

Moreover, with the generation of vast amounts of data that exceed Moore's Law~\cite{csur/Navarro21a, Navarro:2020survey2}, the storage and processing of this data for data science applications have become critical concerns. This is particularly evident in the context of Graph Databases (GraphDBs)~\cite{AnglesG08,LiangMTZ24}, which are emerging as the primary software architecture for applications requiring interconnected data modeling (i.e., data represented as graphs)\footnote{The global Graph Database market was valued at \$2.33 billion in 2023 and is projected to reach \$14.58 billion by 2032~\cite{GD_projection}}. Graphs are essential for network analytics and are increasingly vital in the context of large language models (LLMs) due to the advent of Retrieval-Augmented Generation, which enhances LLMs through the effective use of knowledge graphs \cite{Ji_2022,PengXNO23,9416312,abs-2404-16130}. Efficiently storing and processing large graphs in terms of time and space remains a significant challenge. 

For this reason, sustainability-oriented practices that leverage industrial big data to influence product innovation are gaining importance; the use of industrial big data mediates the relationship between sustainability orientation and product innovation \cite{industrial-big-data}. Furthermore, Green Computing and Green Software Engineering have emerged as responses to sustainability challenges, prompting a paradigm shift that integrates and partially deviates from traditional design, use, and management of computer hardware and software. Recent studies~\cite{ict-emissions} estimate that the ICT sector currently accounts for 2.1-3.9\% of global greenhouse gas (GHG) emissions. Machine learning significantly contributes to these emissions \cite{verdecchia-green-AI}. Energy waste does not affect only data centers, but also small edge and IoT devices that rely on battery life as a critical requirement~\cite{baldini, ladra-green}. Hence, the growing awareness of energy consumption has sparked interest in green computing in the context of AI-enabled IoT infrastructures  (AIoT) too. By ``greening data structures and algorithms,'' software engineers aim to reduce resource utilization, minimize power consumption, and extend battery life through improved algorithm design. However, \cite{programmers-know-energy} observes that programmers often lack experience with software-induced energy consumption, though they increasingly need to consider it a primary constraint when designing their software.

In this paper, we draw attention to space--time efficient and low-carbon footprint representations of large matrices on server machines and resource-constrained devices, such as single-board computers. This problem poses scalability challenges for storage, operation, and bandwidth, both in server-client transmissions and local computations.
For example, in ML applications although training costs can be one-time, or even free with pre-trained models, deploying and running their inference capabilities over an extended period often results in expensive matrix-based computations for client-side RAM and CPU of edge and IoT devices.

Compressed data structures provide a natural solution to mitigate storage occupancy while enabling efficient data transmission and offering opportunities for faster processing. Indeed, computation-friendly compressed representations~\cite{navarro_book} allow for indexing, querying, processing, and manipulating data directly in their compressed format without requiring any prior decompression, thereby saving memory usage and speeding up computations at run time. 

However, despite the critical role of matrix-to-vector computations in scientific computing, efficient machine learning inference, graph analysis, and more, to our knowledge, the only study that examines the actual impact of compressed data structures on energy consumption is that of Fuentes-Sepúlveda and Ladra~\cite{ladra-green}, which focuses on compact integer representations. Given the above discussion, we believe that an analysis of compressed matrix representations from a green viewpoint may provide software engineers with a practical guide that can allow them to select the most energy-efficient matrix format based on the time and space constraints of their problem. As a side outcome, with our paper, we aim to advance the knowledge of the relationship between lossless compression and energy consumption.

\section{Our Contribution}

In this paper, we focus specifically on the space, time, and energy efficiency of matrix-to-vector multiplications using well-known compressed formats for large sparse binary matrices. Sparse matrix-vector multiplication (SpMV) is central to scientific computing applications, as it underpins the solution of sparse linear systems and sparse eigenvalue problems through iterative methods. Additionally, SpMV is essential for graph-analysis algorithms such as PageRank and centrality algorithms and is relevant for inference in binary machine learning models.

The primary research questions we address in this paper are as follows:

\begin{itemize}
\item[Q1] For matrix-vector multiplication, does compression lead to savings in space, time, and energy?
\item[Q2] Are time-optimized matrix formats the most energy-efficient, and vice versa?
\item[Q3] In the context of matrix-vector multiplications, which runtime metrics (such as CPU cycles, number of instructions, and completion time) most significantly impact energy efficiency? 
\end{itemize}

In this study, we focus on the problem of efficient matrix-to-vector multiplications, specifically when the matrix is sparse and Boolean, while the vector is real-valued. Computing the product of a binary adjacency matrix from a large graph with a real-valued vector is a crucial operation in various graph analytics tasks. Among these tasks, we consider the computation of the well-known PageRank score, which assesses the relevance of a Web page (see \Cref{sec:pagerank}). This score has applications beyond the Web domain, and there are extensive datasets available for evaluating performance in real-world scenarios.

We believe that the findings presented in this article will be of interest not only to the Network Analytics community but also to software engineers seeking efficient vector multiplications over binary and ternary matrices like the ones that occur in highly quantized large language models (LLMs). Indeed, binary and ternary LLMs \cite{arxiv-era-1-bit,ternary1, ternary2} offer a favorable trade-off between space and accuracy, and can benefit from compressed binary matrix formats by deploying two matrices to represent the two non-zero values of the quantization. This would be particularly interesting in federated learning~\cite{ternary-federated} because it would enable utilizing resource-constrained nodes such as smartphones or IoT devices, where battery life is a critical concern~\cite{federated-battery}. 

Further, the inference and training stages of Graph Neural Networks (GNNs) are often dominated by the time required to compute lengthy sequences of matrix multiplications involving the sparse graph adjacency matrix and its embeddings. Motivated by this challenge, the authors of \cite{gnn-preprint} recently published preliminary results showcasing efficient matrix multiplication kernels based on new compressed formats for binary matrices. However, their work remains largely in the early stages, and its scalability has yet to be validated, as the authors tested their results solely on adjacency matrices derived from networks of scientific co-authors and citations, which are relatively small in scale. The largest graph they evaluated contains only 540,486 nodes. In contrast, this paper considers matrix formats capable of processing datasets with millions of nodes, even on resource-constrained devices like Raspberry Pis (for instance, our dataset \indochinaX has 7,414,866 vertices).

\smallskip From a technical perspective, we address the problem of compressing Boolean sparse matrices for subsequent fast matrix-to-vector multiplications by extending the results in~\cite{francisco-computation-friendly} by experimentally analyzing three distinct matrix-compression formats: the grammar-based methods described in \cite{mmr-vldb22}, the $k^2$-tree~\cite{k2-tree}, and Zuckerli~\cite{zuckerli}, which is an enhanced version of WebGraph \cite{webgraph}. The appeal of these formats stems from their ability to achieve significant space reduction on disk by exploiting data patterns beyond mere sparsity, while also ensuring that time performance scales with the size of the compressed matrix representation. It is important to note that dense formats requiring quadratic space would impose prohibitive space and time demands, even for matrices containing millions of nodes.

In this paper, we have adapted each compression format to support matrix-to-vector multiplication across multiple threads. For each dataset, we measured the running time and energy consumption of the matrix-to-vector multiplications by executing the PageRank algorithm~\cite{pagerank-original}. We conducted our tests on two platforms: a multi-core \xeon server utilizing up to 48 threads, with energy performance assessed via Intel RAPL~\cite{rapl2}, and a resource-constrained edge device (Raspberry Pi) using up to 8 threads, measuring energy consumption with a benchtop power meter.

We summarize the main contributions of our paper as follows (additional details and contributions are provided in the next sections):

\ignore{Apart from time and space performances, in this paper, we analyze the factors that affect energy consumption (Q1), particularly their impact on lossless matrix compression. Preliminary work~\cite{ladra-green} related to integer vector compression suggests that energy consumption often strongly relates to process completion time, although this is not always the case. Our analysis examines the number of instructions, CPU cycles, and memory loads, highlighting how effective utilization of cache memory affects instructions per clock throughput and, consequently, the time and energy efficiency of data-parallel algorithms as the degree of parallelism increases.\pferragina{e quindi che si trova?}\ftosoni{\cmark Fatto}. To address questions (Q2) and (Q3), we conducted experiments with multi-threaded PageRank computations on an Intel\textregistered\xspace multicore machine, measuring performance using a performance monitoring tool. Our findings indicate that lossless compression can save space, time, and energy for various matrix formats. We also drew conclusions about the relationship between completion time and energy efficiency, as well as how cache efficiency impacts the overall performance of each algorithm. Subsequently, we corroborated our results through laboratory tests on a single-board computer, estimating the energy requirements of each configuration by measuring the current drawn by the device.}

\begin{itemize}

\item We provide an implementation of a multithread matrix-to-vector multiplication algorithm based on the recently introduced Zuckerli compression format which is a tool improving the WebGraph algorithm in terms of compression ratios, yet providing a resource usage for decompression comparable to WebGraph.

\item Compared to \cite{francisco-computation-friendly}, we evaluate a more recent and larger set of {\em computation-friendly} matrix representations. Additionally, we assess the impact of compression on energy consumption, testing not only single-threaded versions as in \cite{francisco-computation-friendly} but also multi-thread implementations.

\item Our analysis involves collecting metrics and comparing results on two different platforms: a multi-core \xeon Gold machine and a single-board Raspberry Pi 4. For the \xeon, we use the Intel RAPL software profiler to gather metrics, while for the Raspberry Pi 4, we utilize a power meter.

\item We examine the energy performance of the PageRank algorithm on both platforms and find that a careful selection of the lossless compression format for the involved matrices can consistently yield energy savings of 1 to 2 orders of magnitude across all tested datasets and both platforms.

\item Concerning question Q1, we observe that although Zuckerli and the $k^2$-tree are more space-efficient, they are generally slower and typically consume about an order of magnitude more energy than the grammar-based solutions presented in \cite{mmr-vldb22}.

\item As for question Q2, we found that the simple model in which energy consumption grows linearly with completion time is not always accurate.  Indeed, we plotted running time and energy consumption in terms of the number of threads and found that in some instances the optimal degree of parallelism for energy efficiency is lower than that for time efficiency, suggesting that it is more difficult to scale energy than to scale time. Hence, in some scenarios, software engineers should consider suboptimal time performance to attain energy-optimal executions, as our results suggest that slightly slower algorithms may result in greater energy savings. 

\item Regarding question Q3, our results indicate that the number of L1 and L3 cache operations is critical for time and energy efficiency in data compression algorithms. Poor utilization of the cache hierarchy can degrade instruction-per-clock-cycle throughput, resulting in increased latencies and energy inefficiencies.

\end{itemize}

The remainder of the paper is structured as follows. In \Cref{sec:related} we discuss related works. In  \Cref{sec:pagerank} we describe the PageRank algorithm and how we implemented it in our experiments. In \Cref{sec:cmr} we discuss compressed matrix representations and we justify the choice we made for our experiments. \Cref{sec:experiments} describes our experimental setup, and \Cref{sec:results} illustrates and discusses the results of our experiments. \Cref{sec:future} summarizes the lessons learned from our analysis and outlines suggestions for future work.

\section{Related Work} \label{sec:related}

\subsection{Green Software Engineering}

As energy has become a leading design constraint for computing devices, hardware engineers and system designers are exploring new models and directions to reduce the energy consumption of their products. However, with green software engineering still in its infancy, most of the efforts in this field have been devoted to hardware-related issues (see, e.g., \cite{review_battery_2010}), with little to no attention on green software~\cite{energy-ale, ladra-green}. 

The work of \cite{energy-ale} advocates for energy management as a guiding principle in application coding and algorithm design, demonstrating that energy-aware hardware and software can reduce energy consumption in legacy applications. They show that increased parallelism not only decreases running time but also reduces energy usage. Their study explores various algorithmic patterns in energy-efficient design, revealing that some patterns yield greater savings; for example, they emphasize the critical importance of parallelizing reads over writes.

The success of ChatGPT has further popularized ML algorithms, resulting in environmental consequences and rising financial expenses~\cite{chatGPT_forbes}. The GPT-3 model, with its 175 billion parameters, exemplifies this problem, as each training iteration costs millions of dollars~\cite{GPT3-expensive}, not to mention the computational costs associated with experimenting with different hyperparameters. Similarly, a 2023 estimate~\cite{wired-gpt4} from OpenAI's CEO placed the training cost for GPT-4 at approximately 100 million dollars, indicating that the future of AI necessitates innovative and more sustainable approaches.

\cite{puzzling-out-sustainability} argued that the definition of Green Software in literature is rather fuzzy and chaotic, as one can find different terms such as green software, green by software, green in software, and the like. While ``green by'' solutions entail IT as a tool to meet sustainability criteria,  ``green in'' techniques concentrate on software and hardware themselves. Green software refers to software for which energy performance can be measured and that promotes energy and resource savings compared to traditional, unoptimized software~\cite{reference-model}. These energy-related algorithmic optimizations, under the more favorable scenarios,  can also reduce the application time and space requirements, thereby improving user experience; or, at least, they attempt to minimize compromise to service.

Oversimplified models for energy complexity, such as the one proposed by \cite{simple-energy-model}, expressed as $T+\left(PB\right)\cdot I$ (where $T$ is the completion time, $I$ is the number of parallel I/Os, $P$ is the number of memory banks, and $B$ is the block size), measure energy consumption asymptotically but fail to capture the constants that reflect the dynamics of real-world machine architectures~\cite{energy-ale}.

\cite{architecting-sustainability} has recently surveyed the experiences and beliefs of IT engineers and practitioners regarding current best practices in architecting for sustainability. They concluded that researchers and practitioners view business motivations and short-term thinking as significant obstacles to genuine sustainability implementation. Both groups also agree that key challenges include a lack of consensus on the importance of sustainability and an absence of knowledge regarding concrete measures that can be taken. Further, \cite{O3-energy} has suggested that power-oriented source code optimizations often do not justify the average programmer's time investment. While source-level power tuning may be effective for tiny embedded programs, it frequently proves challenging for larger production software.

To address the lack of a reference model for assessing software sustainability and the absence of consensus on measurement setups, methods, and techniques for energy data analysis, \cite{reference-model} has published a reference model this year called the Green Software Measurement Model (GSMM) along with a glossary based on the ontology presented by \cite{energy-ontology}. This model, which encapsulates a decade of research by groups from four countries, encourages further contributions from researchers and practitioners within the (Green) Software Engineering community. 

The proposed model emphasizes the significance of power meter devices and advanced software profilers like Intel RAPL for accurate energy profiling estimations. It also underscores the importance of carefully selecting metrics—including CPU and GPU usage, duration, mean power draw, energy consumption, and network traffic—to develop a comprehensive model for profiling an application's energy usage.

Measuring software-induced energy and resource consumption is essential for assessing and mitigating the environmental impact of software. Notable frameworks for measuring energy consumption include the Software Energy and Resource Efficiency Analysis (SERENA), developed at Umwelt-Campus Birkenfeld and based on the work of \cite{SERENA}; the Green Software Measurement Process (GSMP)~\cite{GSMP}; the Sustainability Assessment Framework (SAF) Toolkit\footnote{\url{https://github.com/S2-group/SAF-Toolkit}}; the Green Metrics Tool\footnote{\url{https://github.com/green-coding-solutions/green-metrics-tool}}; the Cloud Energy Usage Estimation Model based on the research of \cite{Interact}; the Software Footprint\footnote{\url{https://www.oeko.de/blog/energieverbrauch-von-software-eine-anleitung-zum-selbermessen/}}; the Emission Estimation Framework from the Sustainable Digital Infrastructure Alliance (SDIA) based on \cite{SDIA}; and the Container Overhead Measurement Methodology introduced in \cite{Kreten2022}. We refer readers to the survey by \cite{reference-model} for a thorough description of these models and their respective application areas.

Notwithstanding these substantial advancements, the pioneering work of \cite{ladra-green} on green compacted data structures argues that among the previous work related to energy efficiency, there is little research within the core area of Algorithmics, with many contributions such as \cite{albers-efficient-algorithms, sigact-algorithmic-power} focussing on energy-saving mechanisms based on speed scaling or power-down mechanisms, where a machine is scheduled to switch between a set of active states (standby, suspend, sleep, full-off), each with its consumption rate.

The increasing size of large language models (LLMs) has posed challenges for deployment and raised concerns about their environmental impact due to high energy consumption. Recent proposals, such as Bitnet~\cite{bitnet}, aim to introduce a 1-bit Transformer architecture specifically designed for LLMs, which can greatly benefit from compact binary formats and optimizations for space, time, and energy in sparse matrix-vector (SpMV) computations. Additionally, LLMs quantized to three values (-1, 0, +1), as demonstrated in \cite{arxiv-era-1-bit}, can also leverage efficient binary matrix formats and SpMV techniques. Ternary matrices can be represented using two distinct binary formats: one for the -1 values and another for the +1 values. The use of binary or ternary formats, as opposed to real-valued ones, allows the replacement of multiplications over FP32 with summations and subtractions between FP32, which are faster in modern computer architectures \cite{Gaurav-survey-DL, ternary2}.

In this paper, we address the challenge of energy conservation from the perspective of green algorithm engineering, focussing on the design of energy-efficient algorithms that minimize energy consumption.

\subsection{Parallel Programming}\label{subsec:parallelism}
It is well-known that parallel programming has great potential for effectively reducing power consumption. Most modern computer systems are based on CMOS (Complementary Metal-Oxide Semiconductor) technology, whose dynamic power consumption varies proportionally to the cube of the voltage. Let's assume a certain software achieves a $1.5\times$ speedup using two cores. A software engineer can exploit this speedup to reduce latency or power~\cite[\S2.5.3]{spm-book}. Assuming the latency requirement is already met, reducing the clock rate by $1.5\times$ can save significant energy. Denoting the energy consumption of the single-threaded application as ${\cal P}_1$ and the data-parallel one as ${\cal P}_2$, the following relation holds

\begin{equation}
{\cal P}_2 = 2 \left( \frac{1}{1.5}\right)^3 \approx 0.6 \: {\cal P}_1
\end{equation}

\noindent where the factor of 2 is due to having two cores. Another common strategy to save energy with parallel computing is to complete the computation as quickly as possible, allowing all threads to be put into a low-power {\em sleep mode}, which consumes much less energy~\cite[\S2.5.3]{spm-book}. The authors of \cite{O3-energy} show that idle or sleep states can yield up to 19\% energy savings and can be easily combined with other energy-saving approaches, also suggesting that increasing parallelization from 1 to 16 threads can save energy for many applications, even the poorly scaling ones. This result is particularly critical for mobile devices, where parallelism can reduce the time the device is powered on (display and other components), yielding a better user experience and reduced energy consumption.

To exploit the potential of parallel computing in green algorithm engineering, we implement and test data-parallel versions of all our code; cf. \Cref{subsec:experiments_parallelism}.

\section{The PageRank Algorithm}\label{sec:pagerank}

We selected the PageRank algorithm~\cite[\S21.2]{IR-book} as an example of an important algorithm involving a sparse and usually highly compressible matrix.  The PageRank algorithm was first used by Google's search engine to evaluate the relevance of Web pages, modeled as vertices in a Web graph. 

Given a directed graph $G = (V, E)$ modeling the Web with $n = |V|$ vertices (pages) and $m = |E|$ edges (links), let $A$ be its adjacency matrix, where $A_{uv} = 1$ if and only if there is a link from vertex $u$ to vertex $v$; $A_{uv} = 0$ otherwise. The normalized adjacency matrix $M$ is defined as $M = D^{-1} \cdot A$, where $D$ is a diagonal matrix whose diagonal elements $d_{u} = \sum_v A_{uv}$ contain the out-degree of each vertex $u$. 
$M$ is the so-called random-walk matrix, where a random walker at vertex $u$ jumps to a neighbor $v$ of $u$ with probability $1/d_{u}$. 

The graph $G$ may have dangling vertices, namely vertices having no outward edges. To avoid a random walker getting stuck in such vertices, {PageRank adds all possible outward edges to all graph vertices. This is equivalent to substituting all empty rows of $M$ with the vector $\vec{1}/n$. This modification ensures that $M$ is a {\em stochastic matrix}, meaning that for each matrix row $i$, it holds that $\sum_{j=1}^nM_{ij} = 1$.} {Moreover, for convergence reasons, PageRank introduces the {\em teleport} step: When reaching a vertex, the random walker selects with some fixed probability (say $0 < \alpha < 1$) whether it has to continue by traversing a graph edge or jump to a random vertex in the graph.} In their original paper~\cite{pagerank-original} Brin and Page suggest setting $\alpha = 0.15$. With the above notation, the PageRank vector giving the relevance of each page is computed by the unique limit of the sequence~\cite{pagerank-recursion}

\begin{equation}\label{eq:iterative}
\vec{\pi}_{t} = 
\alpha\cdot \vec{\pi}_{0} + (1 - \alpha) \cdot \vec{\pi}_{t-1} \cdot M
\end{equation}
In \cite{pagerank-original}, the initial probability distribution $\vec{\pi}_0$ is set to $\vec{1}/n$; however, it can be any non-zero vector and still guarantee convergence to the same PageRank vector.

\subsection{Implementation details}

In PageRank implementations, the iteration~\ref{eq:iterative} is repeated until the difference between two consecutive estimates $|\vec{\pi}_{t}-\vec{\pi}_{t-1}|$ goes below an assigned threshold. The convergence of the method is guaranteed by the mathematical properties of the iteration matrix. In practice, we do not explicitly build the matrix $M=D^{-1} A$ but, instead, we operate with the adjacency matrix $A$ of the input graph. The diagonal matrix $D$ is represented by an array $O[1,n]$, where the entry $O[u]$ represents the out-degree $d_u$ of vertex $u$.

Formula~\ref{eq:iterative} involves left products, but since most compressed matrix representations are row-oriented, meaning they represent matrices in row-major order, it is a standard practice~\cite{francisco-computation-friendly} to transpose formula~\ref{eq:iterative} so that the left matrix-vector product becomes a right product involving $M^t = A^t D^{-1}$. Note that $\left(D^{-1}\right)^t=D^{-1}$, since $D$ is a diagonal matrix.

\section{Compressed Matrix Representations}\label{sec:cmr}

\subsection{WebGraph and Zuckerli}\label{subsec:zuckerli}

Many (binary) graphs extracted from the Web contain redundancies and dependencies that can be exploited by proper compression tools to reduce the storage size of these graphs. The first ad-hoc compression method for Web-related graphs, WebGraph~\cite{webgraph}, was proposed over 20 years ago. It effectively compresses adjacency lists by leveraging the similarities between lists of close-by vertices within a graph. As an example, Web pages within the same domain often share outgoing links to a common set of destination pages. Therefore, such common links can be represented once in a list from which subsequent lists can be copied. WebGraph exploits this copying property to compress each adjacency list based on a reference list. Although it was originally implemented in Java, it has since been translated into C++~\cite{webgraph++}; lately, WebGraphs's original authors also published a Rust implementation~\cite {webgraph-rust}.

Zuckerli \cite{zuckerli} is a more recent compression method for Web graphs that applies advanced compression techniques and novel heuristics on top of WebGraph, further improving its compression. Zuckerli provides both a compressed representation based on the Asymmetrical Numeral Systems (ANS)~\cite{ans_moffat} encoding for storage and one based on Huffman encoding~\cite[\S12.1]{pearls_ferragina}\cite[\S16.3]{cormen_leiserson_book}, allowing fast direct access to the individual adjacency lists of the compressed graph without decompressing it in its entirety.

The authors of \cite{francisco-computation-friendly} observed that these graph representations provide a computation-friendly framework for efficient right matrix-to-vector multiplications. {The main idea is to exploit the copy-property of adjacency lists between vertex $v_i$ and its reference $v_{r_i}$ so that it is easy to evaluate $v_i \cdot x$ by simply copying the result of $v_{r_i} \cdot x$ and then summing up the contributions of the differences between the adjacency lists of $v_{i}$ and $v_{r_i}$.} We build upon this idea to carry out matrix-to-vector multiplications on top of Zuckerli-compressed matrices. We used the code from Google's GitHub repository \url{https://github.com/google/zuckerli} and Zuckerli's default compression settings.

\subsection{k2-tree}\label{subsec:k2tree}

The $k^2$-tree \cite{k2-tree}\cite[\S9.2.1]{navarro_book} is a compact data structure designed to compress adjacency matrices of size $n^2$, by exploiting the sparsity and clustering of $1$'s. The data structure underneath the $k^2$-tree is a $k^2$-ary tree where each vertex represents a submatrix of the adjacency matrix. For simplicity, the size is zero-padded to the next power of $k^2$. The tree root represents the whole matrix and each of its children represents a submatrix of size $n^2/k^2$. If a submatrix is empty (i.e., it contains only $0$'s), the corresponding vertex is represented by a bit set to 0. In all other cases, the corresponding vertex is represented by a bit set to 1. Once the first level is built, the procedure continues recursively only into the submatrices with $1$'s, by splitting each of them into $k^2$ smaller submatrices. Thus, the height of a $k^2$-tree is always $1+\log_k n$, independently of the specific distribution of $1$'s and $0$'s within the input matrix; the access to a single cell costs $\Oh \left(\log_k n\right)$. Note that a submatrix is not split further when full of $0$'s.

\begin{figure*}[t]
\centering
\hfill
\subcaptionbox{}{\includestandalone[height=0.25\textwidth,mode=buildnew]{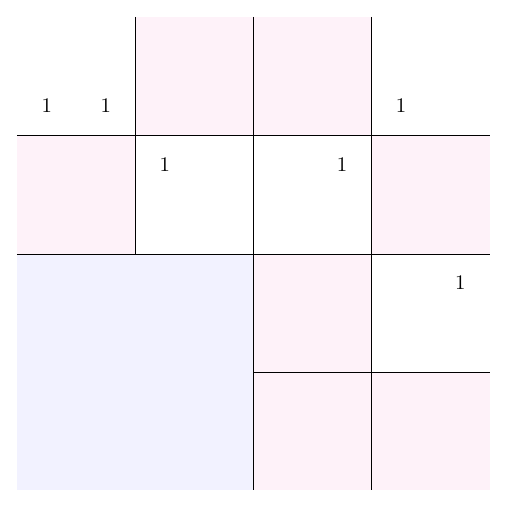}}%
\hfill
\subcaptionbox{}{\includestandalone[height=0.25\textwidth,mode=buildnew]{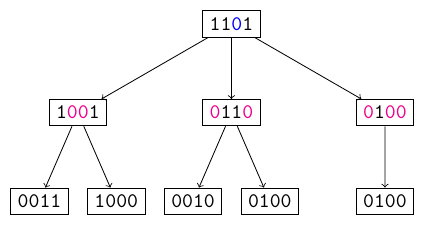}}%
\hfill
\caption{A 8x8 binary matrix (a) and the corresponding $k^2$-tree (b).}
\label{fig:k2-tree}
\end{figure*}

Figure~\ref{fig:k2-tree} depicts the $k^2$-tree ($k=2$) for a $8\times8$ binary matrix. The arrows in the figure are included just for the sake of presentation, but the actual implementation of the $k^2$-tree is pointer-free because it consists of a mere sequence of bits. The $k^2$-tree represents the submatrices in row-major order; lower levels of the $k^2$-tree correspond to smaller submatrices, with the root representing the original matrix and the leaves representing single cells. 

Recently, \cite{k2-tree_brisaboa, k2-tree_spire2023} have shown that the $k^2$-tree allows for carrying out matrix-to-matrix sums and multiplications in the compressed domain, yielding the resulting matrix directly in its compressed $k^2$-tree representation and without the need of constructing {\tt rank} data structures. In this paper, we are interested in right matrix-to-vector multiplications, for which we do not need {\tt rank} support over the bit vectors representing the $k^2$-tree; indeed, a simple visit to each tree node performed in any order is enough to implement the multiplications underneath our PageRank calculations.

For our experiments, we run the C++ implementation of the $k^2$-tree made available on the website of the University of A Coruña~\cite{k2-tree_udc} at \url{https://lbd.udc.es/research/k2tree/}. Other implementations have been made available in the SDSL library~\cite{sdsl} at \url{https://github.com/simongog/sdsl-lite/blob/master/include/sdsl/k2_tree.hpp} or, more recently, in the codebase of \cite{k2-tree_spire2023} at \url{https://github.com/adriangbrandon/rpq-matrix/tree/main/lib/matrix_gn}. We have experimented with all those and found that these latter two did {\em not} outperform in space, time, and energy the one implemented at A Coruña on our tested datasets; thus we finally resolved to use the version of A Coruña in our experiments. We however modified this version by removing all {\em rank}-related data structures, as we do not need them for the multiplications, thus using slightly less space than the implementation in~\cite{k2-tree_udc}.

\subsection{RePair compressed matrices}\label{subsec:mm-repair}

The technique of matrix multiplications on RePair compressed matrices (mm-repair, for short) has been introduced in \cite{mmr-vldb22}; it compactly represents the non-zero entries of a matrix using a grammar that captures regularities between rows, reducing the representation size {\em and} speeding-up matrix operations. 

\begin{figure*}[t]
\centering
\hfill
\subcaptionbox{}{\includestandalone[width=0.3\linewidth,mode=buildnew]{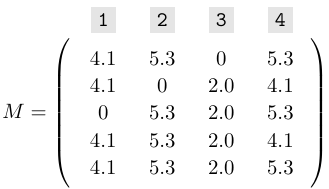}}%
\hfill
\subcaptionbox{}{
\centering
  \includestandalone[width=.24\linewidth, mode=buildnew]{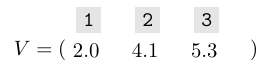}
  \includestandalone[width=.4\linewidth,mode=buildnew]{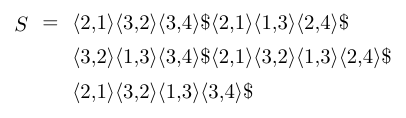}
}
\caption{A matrix $M$ (a) and the corresponding $(V,S)$ representation (b).}
\label{fig:mmr_vs_ss}
\end{figure*}

Given a matrix $M$, the scheme firstly produces a representation consisting of a dictionary $V$ of distinct values and a sequence $S$ containing column-value pairs, as shown in Figure~\ref{fig:mmr_vs_ss}. $S$ is obtained by scanning the matrix $M$ row-by-row: each non-zero entry $m_{ij}$ in $M$ is represented in $S$ by the pair $\sy{\ell}{j}$, where $\ell$ is the index of the value $m_{ij}$ in $V$. At the end of each row, a unique delimiter $\sz{i}$ is appended to~$S$. Note that the above scheme works for any real-valued matrix $M$, though in this paper the input matrices are binary; thus, the dictionary $V$ only contains the value~$1$.

\begin{figure}[t]
\centering

  \includestandalone[width=.8\linewidth, mode=buildnew]{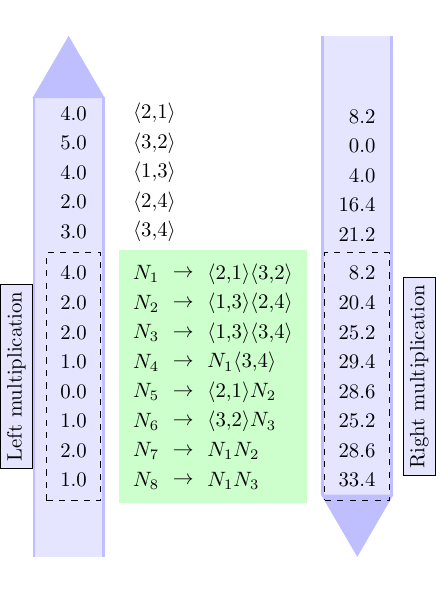}

  \includestandalone[width=.6\linewidth, mode=buildnew]{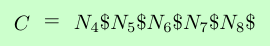}

\caption{The MMR-compressed $M$ of Figure~\ref{fig:mmr_vs_ss}.}
\label{fig:mmr_rs_cs}
\end{figure}

The sequence $S$ is then compressed using RePair~\cite{larsson2000off} that computes a grammar representing $S$. RePair's output consists of the rules set $R$ and a final sequence $C$ of non-terminals interleaved with the row delimiter $\sz{i}$. For our example matrix $M$, the set $R$ (eight rules of the form $N_i \to A_i B_i$) and the sequence $C$ are shown in Figure~\ref{fig:mmr_rs_cs} with a green background. The representation of $M$ thus consists of the triplet $\left( R, C, V \right)$. We now quickly review how this representation supports right and left matrix-to-vector multiplications, without the need for decompressing the entire matrix, taking time $O(\lvert R \rvert + \lvert C \rvert)$ and using $O(\lvert R \rvert)$ additional space. For the underlying theory and a thorough explanation of this scheme, the reader is referred to the original article \cite{mmr-vldb22}.

For right multiplications $y = Mx$, we perform a single scan over the rules $R$ of the grammar from the first grammar rule up to the last one and evaluate the contribution of each rule given the values of the vector $x$. To evaluate a terminal symbol $\sy{\ell}{j}$, we perform the product $\Val_x(\sy{\ell}{j}) = V[\ell] \cdot x[j]$. For each nonterminal symbol $N_i$ appearing on the left-hand side of rule $N_i \to A_iB_i$, we evaluate the sum $\Val_x(N_i) = \Val_x(A_i) + \Val_x(B_i)$. We store the results of the evaluated nonterminals in an auxiliary array $W$ (dashed in Figure~\ref{fig:mmr_rs_cs}) whose size is $\Theta(\lvert R \rvert )$. The components of the resulting vector $y$ are eventually determined by evaluating the nonterminals appearing in $C$. Figure~\ref{fig:mmr_rs_cs} illustrates the procedure to right-multiply $M$ by the vector $x=(2.0, 0.0, 2.0, 4.0)$, with the values $W$ in the blue arrow corresponding to the evaluations of terminals and nonterminals of the grammar.

To carry out a left multiplication $y^t = z^t M$, we follow a dual procedure scanning the rules backward. The left-hand side of Figure~\ref{fig:mmr_rs_cs} schematizes the left multiplication $z^T \cdot M$, with $z^T = \left( 1.0, 0.0, 1.0, 2.0, 1.0 \right)$. To begin with, if a non-terminal $N_i$ appears in position $j$ of $C$, we initialize $W[i] = z[j]$. Then we scan the rules bottom-up and for each rule $N_i \to A_iB_i$ we increase the entries corresponding to $A_i$ and $B_i$ by the value of $W[i]$. Eventually, for each non-terminal symbol $\sy{\ell}{j}$ appearing on the right-hand side of the $i$th rule, we increase the zero-initialized component $y[\ell]$ of the product vector $y$ by the quantity $V[j]\cdot W[i]$.

Experiments in \cite{mmr-vldb22} have shown that $R$ and $C$ can be further compressed resulting in more compact representations that require the on-the-fly decompression of $R$ and $C$. In this paper, we have experimented with the following three variants:
\begin{description}
    \item[Re32:] $R$ and $C$ are no further compressed and are represented as 32-bit integers; this is the fastest variant.
    \item[Reiv:] $R$ and $C$ are represented as packed arrays, with entries of $1+ \log_2 N_{\max}$ bits ($N_{\max}$ being the largest value in $R$ or $C$). In our implementation, we used the class {\tt int\_vector} from the sdsl-lite library~\cite{sdsl}.
    \item[Reans:] $R$ is represented with a packed array and $C$ is represented with the {\tt ans-fold} entropy encoder from~\cite{ans_moffat}; this is the variant achieving the maximum compression. 
\end{description}

\subsection{Other Compressed Matrix Formats}\label{sec:other}

RePair-compressed matrices can be regarded as a generalization of the Compressed Linear Algebra (CLA) system~\cite{elgohary2019compressed} which uses simpler compression schemes. We did not test CLA since the experiments in~\cite{mmr-vldb22} have shown that matrix multiplication with RePair-based algorithms is faster and uses much less memory than CLA, especially for highly compressible matrices.

Another matrix compression scheme that supports matrix-vector multiplication in time proportional to the compressed graph size is the bicliques representation~\cite{extract_bicliques}. A biclique is a pair of vertex sets $\langle S, T \rangle$ such that every vertex in $S$ is connected to every vertex in $T$. Thus a list of $|S|+|T|$ vertex id's is sufficient to encode all $|S| \cdot |T|$ edges. The edges not belonging to a biclique are compressed separately, for example employing a $k^2$-tree. With a proper algorithm for extracting bicliques, the resulting representation can be very space efficient for Web and social graphs, and it can be adapted to support matrix-vector multiplication~\cite{francisco-computation-friendly}. We did not test this approach since only a proof-of-concept implementation is available.

Another recently introduced compressed representation of binary matrices is the Two-Dimensional Block Tree (2DBT) \cite{2DBT} that combines the idea of recursive subdivisions of the $k^2$-tree with copying redundant portions of the input, a feature introduced by the Block Tree \cite{block_trees}. At its core, the 2DBT is an enhanced $k^2$-tree that includes an option to represent submatrices containing a single $1$ in constant space and an option to store a submatrix (called \textit{target}) as a copy of a previous submatrix of the same size (called \textit{source}). As for the space, the 2DBT has the potential to outperform the $k^2$-tree; however, the problem of matrix-vector multiplication with this matrix representation has not been considered to date. Indeed, the design of a multiplication algorithm running in time proportional to the size of the 2DBT compressed representation appears to be a non-trivial problem. For this reason, we resolved not to test 2DBT in our experiments.

\section{Experimental setup}\label{sec:experiments}

\subsection{Transparency and Reproducibility}
The entire codebase to reproduce the experiments of our paper is made available at \url{https://gitlab.com/ftosoni/green-lossless-spmv}.

\subsection{Datasets}\label{sec:datasets}

\begin{table}[ht]
\caption{\textbf{Web graph datasets}}
\label{table}
\centering
\begin{tabular}{rrrc}
\toprule
Dataset & \#vertices & \#edges & Web graph?\\
\midrule
\euX & 862 664 & 19 235 140 & \cmark \\ 
\hollywoodX & 1 139 905 & 57 515 616 & \xmark \\
\inX & 1 382 908 & 16 917 053 & \cmark \\ 
\ljournalX & 5 363 260 & 79 023 142 & \xmark \\
\indochinaX & 7 414 866 & 194 109 311 & \cmark \\

\ukX & 18 520 486 & 298 113 762 &\cmark \\
\arabicX & 22 744 080 & 639 999 458 &\cmark \\
\ukXX & 39 459 925 & 936 364 282 &\cmark \\
\itX & 41 291 594 & 1 150 725 436 &\cmark \\
\bottomrule
\end{tabular}
\label{tab:webgraphs}
\end{table}

We used different real datasets from WebGraph framework \cite{datasets1,datasets2}, available from \url{https://sparse.tamu.edu/LAW}; \Cref{tab:webgraphs} reports their specifications. As shown in the last column, most graphs are Web graphs, in that they derive from crawlings of one or more Web domains. The vertices in these graphs follow the lexicographical order of their reversed URL. It has been empirically observed this helps compression since adjacent rows usually have very similar 0/1 patterns. Though the PageRank algorithm has been designed for Web graphs, we have also included two social network graphs to measure the effectiveness of the compression algorithms in different settings. \hollywoodX is a graph of movie actors. Vertices are actors, and two actors are joined by an edge whenever they appear in a movie together: this is the only symmetric graph in the collection. \ljournalX is a directed graph described in \cite{ljournal-origin} and representing friendships in the social network LiveJournal. In this social network, the notion of friendship is asymmetric, and thus the graph is directed.

\begin{table*}[ht]
\caption{\textbf{Disk occupancy}}
\label{tab:disk}
\centering
\begin{tabular}{r||rrr|r|r||rr}
\toprule
dataset & \reint{} & \reiv{} & \reans{} & $k^2$-tree & Zuckerli & {\tt gzip} & {\tt xz} \\
\midrule

eu-2005 &9.56 &7.58 &6.76 &4.07 &2.26 &2.05 &0.46 \\
hollywood-2009 &14.17 &11.26 &10.52 &7.22 &4.22 &2.18 &0.87 \\
in-2004 &11.38 &8.98 &7.26 &2.92 &1.31 &1.93 &0.39 \\
ljournal-2008 &26.99 &21.34 &19.42 &14.27 &9.69 &2.71 &1.52 \\
indochina-2004 &5.86 &4.93 &4.09 &2.41 &0.79 &1.87 &0.27 \\
uk-2002 &9.34 &8.23 &6.67 &3.12 &1.29 &1.94 &0.40 \\
arabic-2005 &6.20 &5.49 &4.64 &2.75 &0.96 &1.89 &0.32 \\
uk-2005 &6.79 &6.12 &5.03 &2.72 &0.96 &1.90 &0.34 \\
it-2004 &6.16 &5.56 &4.64 &2.76 &0.97 &1.89 &0.32 \\

\bottomrule
\end{tabular}
\end{table*}

\subsection{Computer Architecture}\label{subsec:archi}
  \begin{figure*}[t]
    \centering
    \includestandalone[width=\textwidth,mode=buildnew]{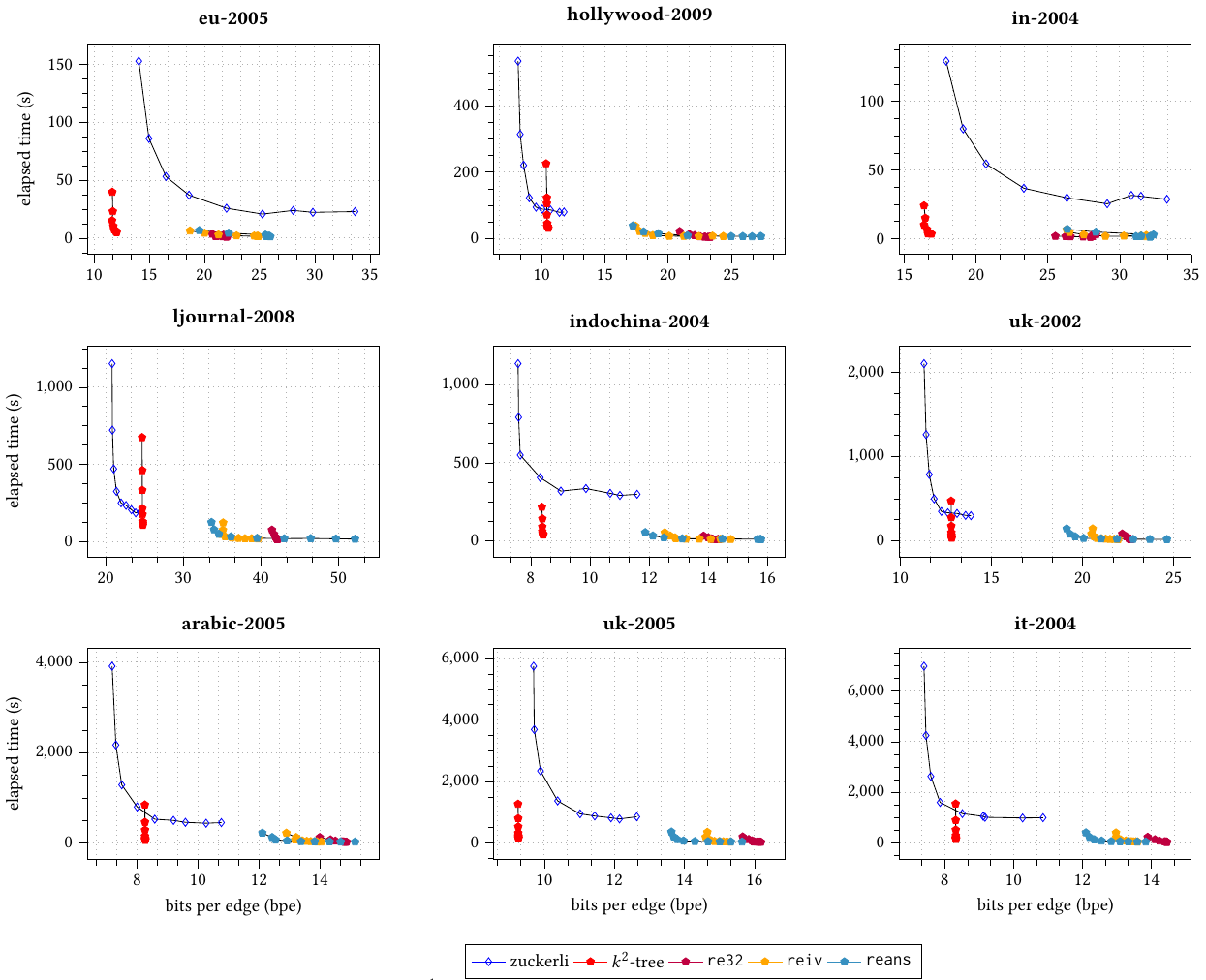}
    \caption{Bits per edge (x-axis) and elapsed times (y-axis) for 100 iterations of PageRank on the \xeon Gold for different numbers of threads.}
    \label{fig:xeon_bpe-elapsed}
  \end{figure*}

We compiled and executed our codes on a Non-Uniform Memory Access (NUMA) server machine equipped with 2 sockets, each having 14 \xeon Gold 6132 CPU @ 2.60GHz, for a total of 28 physical cores. The logical cores are 56 due to Intel\textregistered\xspace Hyper-Threading. The machine is equipped with 12 memory modules of DDR4 RAM, each of them having a capacity of 32 GB for a total of $384$ GB of installed memory. The memory modules work at a velocity of 2666 MT/s. The machine runs Ubuntu 22.04.3 Long-term support (LTS) in 64-bit mode. Each of the 28 cores has L1d (896 KiB), L1i (896 KiB), and L2 (28 MiB) caches. Further, each of the two sockets has an L3 cache, whose size is 38.5 MiB. As for the cache associativity of L1 and L2 caches, which have become a major source of energy consumption~\cite{associative-L1-expensive}, our L1d and L1i caches are 8-way set associative; the L2 caches are 16-way set associative; and the L3 caches are 11-way set associative.

We repeated the experiment on a resource-constrained single-board computer: a Raspberry Pi 4 model B, equipped with four ARM Cortex-A72 CPUs clocked at 1.5GHz. Each of the four cores has its own L1d and L1i cache memories (128+192 KiB); the single shared L2 cache has a size of 1 MB. L1d, L1i, and L2 cache memories are 2-, 3- and 16-way set associative. The device has 4 GB of Low Power Double Data Rate 4 Synchronous Dynamic Random-Access Memory (LPDDR4 SDRAM). The machine runs a 64-bit Ubuntu Server 24.04 LTS. For the experiment on this device, we used the datasets with less than $10^7$ vertices (cf. \Cref{tab:webgraphs}), namely \euX, \hollywoodX, \inX, \ljournalX, and \indochinaX.

\subsection{Code Setup}
As a preprocessing step for our experiments, we extracted the adjacency matrix $A$ from each dataset, transposed it, and compressed it via Zuckerli, the $k^2$-tree and mm-repair (see \Cref{sec:cmr}). We also precomputed and stored on disk the array $O[1,n]$ of the out-degree for each vertex.
For each compression format and each input matrix, we executed 100 iterations of PageRank.

All tested codes were written in C/C++ and compiled via the flag {\tt -O3}, which enables the highest level of optimization that the compiler can perform, including function inlining, loop unrolling, and Single Instruction Multiple Data (SIMD) instructions.  The authors of \cite{O3-energy} showed that GCC's {\tt -O3} option offers significant energy savings, as {\tt -O3} optimized software consumes less than 43\% of {\tt -O0} optimized.

\subsection{Data Parallelism}\label{subsec:experiments_parallelism}
In light of the motivations presented in \Cref{subsec:parallelism}, we implemented and tested a data-parallel solution for all the tested matrix formats. We divided each matrix into as many row blocks as the number of threads. We compressed each block using the tested matrix formats and then ran a data-parallel PageRank on top of the compressed blocks. We followed this approach since right matrix-to-vector multiplications are {\em embarrassingly parallel}, meaning they do not require any specific synchronization data structure beyond a simple barrier to ensure each thread has completed its assigned job. For each algorithm, we forked and joined different threads in C/C++ using standard mechanisms based on POSIX Threads (Pthreads).

\subsection{Power Measurement}\label{subsec:power}

As in \cite{ladra-green}, for the \xeon, we measured the power consumption of our server machine during PageRank calculations by using the Intel RAPL (Running Average Power Limit) interface \cite{intel64-ia32-vol3b}. RAPL works by aggregating the values of several specialized Model Specific Registers (MSRs) to estimate the energy consumption at the core level (all cores in a processor), package level (all cores, memory controller, last-level cache, and other components), and DRAM memory level. Depending on the processor model (in our case, \xeon), one may have access to the energy estimation at the core and package levels, which is the case in our experiments. 

Consistent with \cite{ladra-green}, we present the energy estimation at the package level. Previous studies on energy consumption \cite{rapl1, rapl2} have demonstrated that RAPL provides reasonably accurate measurements. RAPL is central to the Scaphandre measurement and visualization tool\footnote{\url{https://web.archive.org/web/20230623133912/https://01.org/blogs/2014/running-average-power-limit-–-rapl}} 
and is also utilized by CodeCarbon\footnote{\url{https://github.com/mlco2/codecarbon}} and the Experiment-Impact Tracker \cite{henderson2020towards}. Additionally, it is integrated into Green Lab~\cite{greenlab} and the Green Metrics Tool (GMT), developed by Green Coding Berlin, as cited in \cite{reference-model}.

We used the Linux profiler \perf (version 5.15.149) to obtain the Intel RAPL energy estimations. \perf also allows us to measure additional metrics, such as CPU cycles, cache hits and misses of the L1 and last-level caches, and the number of instructions. 

In our experiments with the single-board CPU, we measured the current drawn during PageRank computations by connecting a Fluke 8845A benchtop multimeter in series with the USB-C power cable of the board. The multimeter was configured to a range of 10 A, with a resolution of 4\textonehalf\xspace digits, and a sampling frequency of 3 Hz. Taking into account the frequency spectrum of the current signal, we believe that this sampling rate is adequate for accurately reconstructing the signal with high fidelity.

\subsection{Collected Metrics}

\Cref{tab:disk} tabulates the disk occupancy for the single-threaded matrix formats as bits per edge; the two rightmost columns report, for the sake of comparison, the space on disk for the same matrix compressed using {\tt gzip} and {\tt xz} in their default setting: we compressed using either a 32-bit representation of the edge descriptions for each graph, consisting of sourceId-destinationId pairs. One can see that, in terms of disk occupancy, \reint{} > \reiv{} > \reans{} > $k^2$-tree > Zuckerli. {\tt gzip} is sometimes better and other times worse than 
Zuckerli, and {\tt xz} is the most succinct representation. However, we stress the fact that experiments on matrix-vector multiplications in \cite{mmr-vldb22} showed that {\tt gzip} is at least an order of magnitude slower than algorithm \reint{}; we did not test multiplications based on on-the-fly decompression for {\tt xz}-compressed matrices since {\tt xz} is even slower in decompression than {\tt gzip}. For these reasons, we did not test Pagerank on {\tt gzip}- or {\tt xz}-compressed matrices.

In all our experiments we measured the peak memory usage (PMU) for PageRank executions using the command-line {\tt time} utility and we report it using again the number of bits per edge, as in \Cref{tab:disk}. The PMU will be consistently larger than the disk usage reported in \Cref{tab:disk} for two factors: the first one is that the execution of the algorithm requires additional temporary data structures, and the second one is that splitting the matrix into blocks, to support multithread multiplication, usually reduces the overall compression.

\begin{figure*}
    \captionsetup[subfigure]{labelformat=empty}
    \centering

    \begin{subfigure}[t]{0.32\textwidth}
        \caption{eu-2005}
        \includegraphics[width=\textwidth]{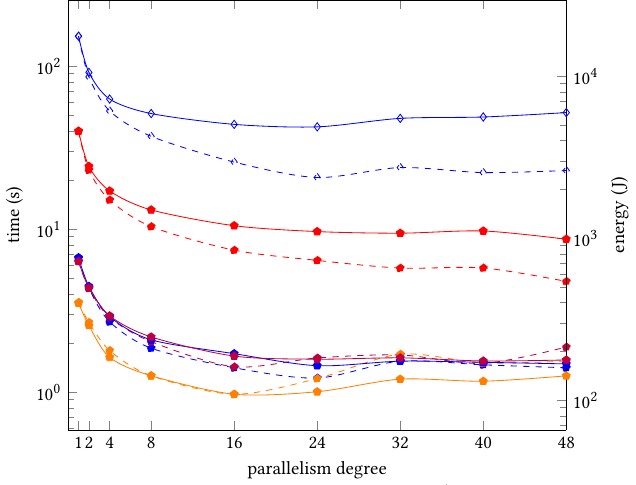}
    \end{subfigure}
    \hfill
    \begin{subfigure}[t]{0.32\textwidth}
        \caption{hollywood-2009}
        \includegraphics[width=\textwidth]{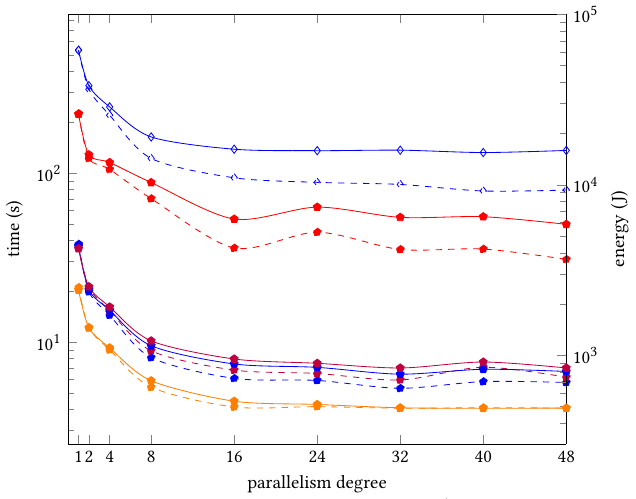}
    \end{subfigure}
    \hfill
    \begin{subfigure}[t]{0.32\textwidth}
        \caption{in-2004}
        \includegraphics[width=\textwidth]{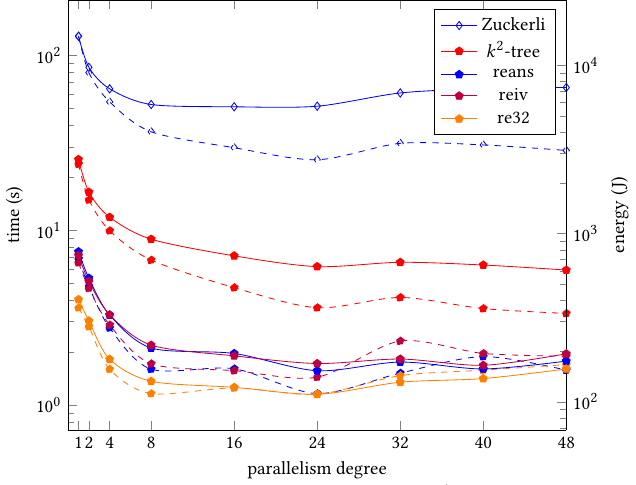}
    \end{subfigure}

    \vspace{1em} 

    \begin{subfigure}[t]{0.32\textwidth}
        \caption{ljournal-2008}
        \includegraphics[width=\textwidth]{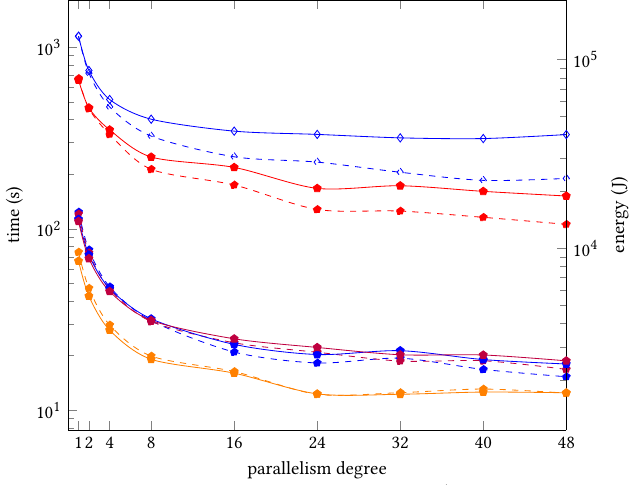}
    \end{subfigure}
    \hfill
    \begin{subfigure}[t]{0.32\textwidth}
        \caption{indochina-2004}
        \includegraphics[width=\textwidth]{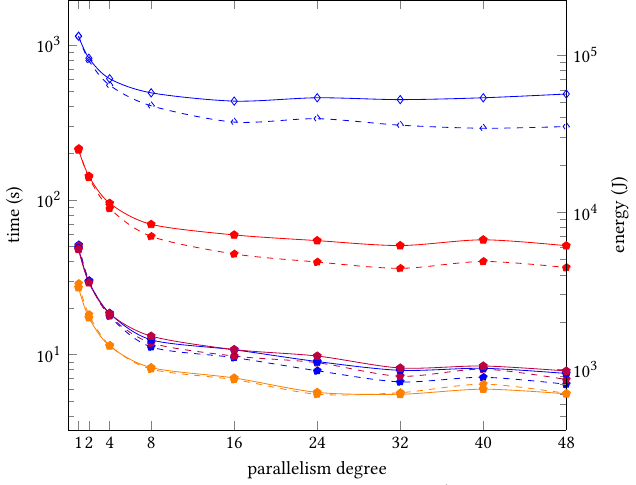}
    \end{subfigure}
    \hfill
    \begin{subfigure}[t]{0.32\textwidth}
        \caption{uk-2002}
        \includegraphics[width=\textwidth]{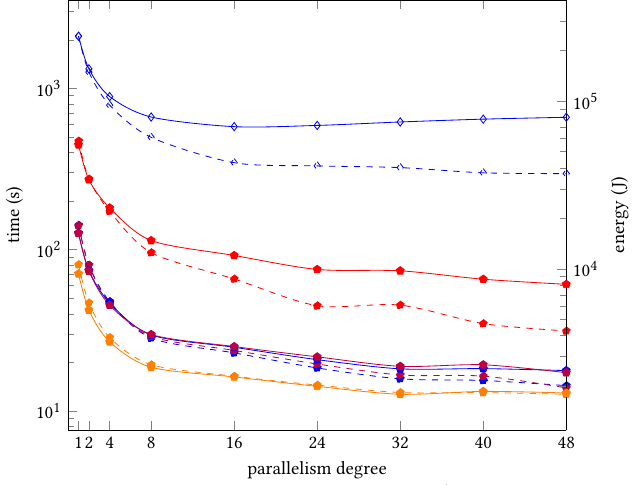}
    \end{subfigure}

    \vspace{1em} 

    \begin{subfigure}[t]{0.32\textwidth}
        \caption{arabic-2005}
        \includegraphics[width=\textwidth]{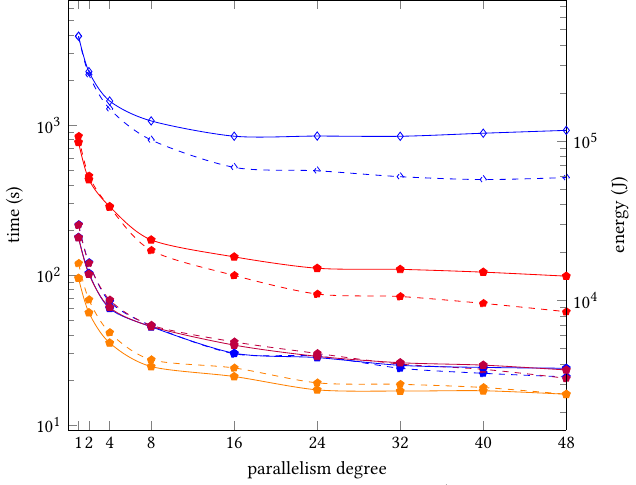}
    \end{subfigure}
    \hfill
    \begin{subfigure}[t]{0.32\textwidth}
        \caption{uk-2005}
        \includegraphics[width=\textwidth]{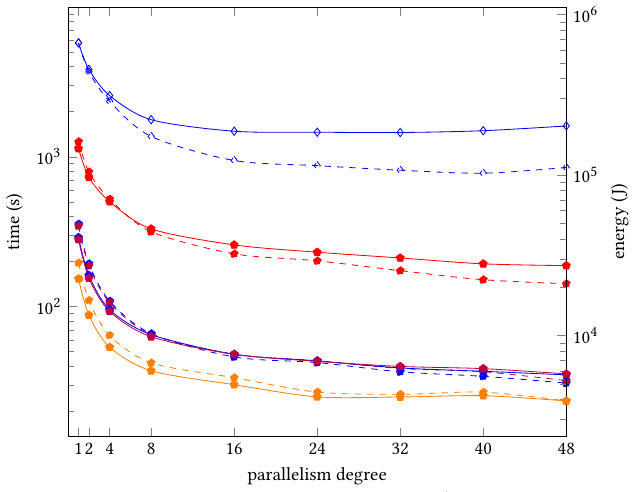}
    \end{subfigure}
    \hfill
    \begin{subfigure}[t]{0.32\textwidth}
        \caption{it-2004}
        \includegraphics[width=\textwidth]{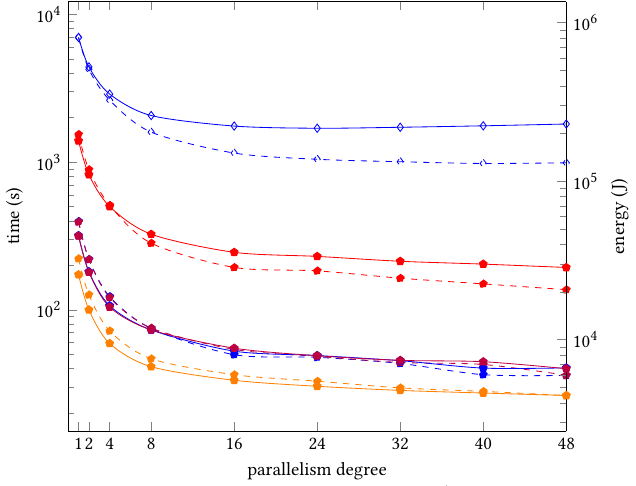}
    \end{subfigure}

\vspace{0.6cm}
\caption{Time performance in seconds (dashed line; left y-axis) and energy usage in Joules (solid line; right y-axis) on the \xeon as a function of the number of threads (x-axis).}\label{fig:optimality_xeon}
\end{figure*}

\section{Results and Discussion}\label{sec:results}

\subsection{Experiments on the Intel Xeon Server}

\Cref{fig:xeon_bpe-elapsed} shows the total elapsed time for 100 iterations of PageRank as a function of the peak memory usage (PMU) in bits per edge. We use markers to differentiate algorithms. Each subfigure represents a different dataset and shows nine data points for each algorithm representing executions with 1, 2, 4, 8, 16, 24, 32, 40, and 48 threads. 

We observe that Zuckerli is generally the most compressed but the slowest solution, while \reint is the fastest, albeit more space-consuming. When using a single thread, the $k^2$-tree is often more than $3\times$ faster than Zuckerli; for the datasets \euX, \inX, and \ukXX, the $k^2$-tree outperforms Zuckerli across all degrees of parallelism, resulting in approximately $3\times$ more lightweight in space. For the remaining six datasets, the single-threaded $k^2$-tree requires $2\times$ more disk space than Zuckerli which, however, degrades significantly faster with increasing parallelism. We believe this is due to Zuckerli’s more complex compression strategy, which involves a sophisticated method for selecting the reference list to represent each adjacency list. Indeed, as the degree of parallelism increases, the dimension of the adjacency-matrix row block assigned to each single thread decreases, leading to a reduction in the number of candidate reference lists for each adjacency list.

The single-threaded grammar-based solutions \reint, \reiv, and \reans provide faster alternatives, but they are approximately $1.5\times$ more space-consuming than Zuckerli and the $k^2$-tree. Recall that the grammar-based compressors can in principle handle any input matrix, that is, they are not specialized to binary matrices such as Zuckerli and the $k^2$-tree. In addition, the grammar-based algorithms also support efficient left matrix-to-vector multiplication, which though not needed in our PageRank implementation, is essential for instance in conjugate gradient methods~\cite{mmr-vldb22}. Among the grammar-based algorithms, \reans is often the most space-efficient. However, as the number of threads increases, its space requirements increase significantly compared to \reint and \reiv, which are more robust and less sensitive to increases in thread count.

In our green computing scenario, we are particularly interested in the energy consumption of the different algorithms. \Cref{fig:optimality_xeon} compares the time (dashed lines) and energy (solid lines)  performances of PageRank computations, on a semilogarithmic scale, as a function of the number of threads. As anticipated, performance improves with increasing degrees of parallelism until the optimal level is reached. Beyond this point, no further speedup occurs, and the algorithms may even experience slower execution times. We immediately observe that increasing the number of threads beyond the optimal parallelism degree for time often results in worse energy efficiency. In examining the energy performances reported in the same graph, it becomes clear that energy does not scale as well as time. Energy savings are generally more limited than time savings, and the optimal parallelism degree for energy is usually lower than that for time. This trend is consistently evident, particularly for the slower algorithms, namely Zuckerli and the $k^2$-tree.
For instance, for \hollywoodX, time performance keeps slightly improving up to at least 40 threads, while energy consumption does not decrease beyond 16 threads. In the cases of \inX and Zuckerli, we find an optimal parallelism degree for the time performances for 24 threads, while energy consumption does not decline past 8 threads.

This observation challenges the traditional rule of thumb that energy optimization strategies are equivalent to time optimization strategies. Rather, our experiments suggest that software developers of multi-threaded scientific computing should adopt a multicriteria approach, taking into account both time and energy performance. For example, they could select the fastest solution that meets specific energy consumption constraints.

  \begin{figure*}[t]
    \centering
    \includestandalone[width=\textwidth,mode=buildnew]{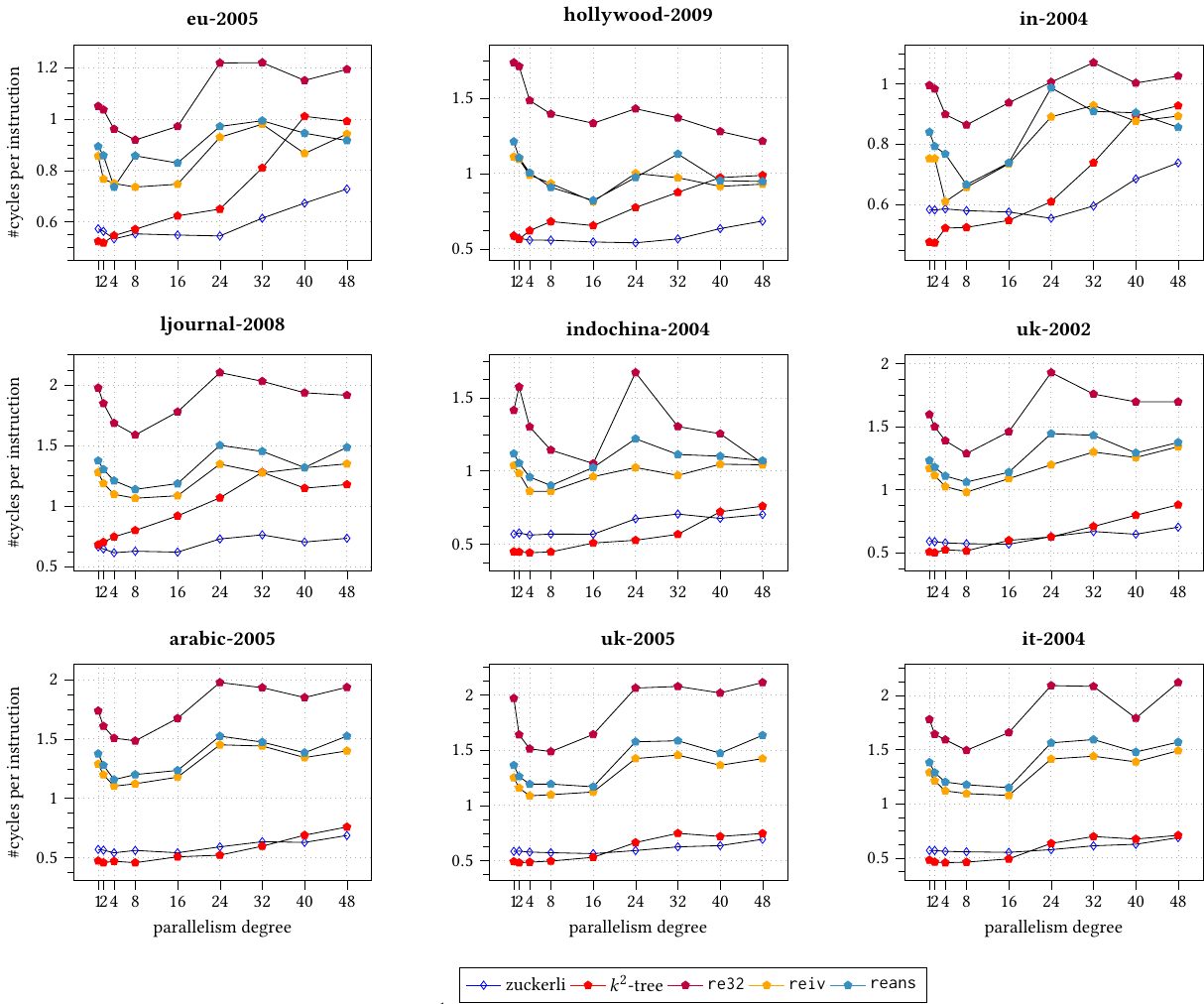}
    \caption{Number of cycles per instruction (y-axis) as a function of the parallelism degree (x-axis) for 100 iterations of PageRank on the \xeon Gold.}
    \label{fig:xeon_pardegree-cyclesinstr}
  \end{figure*}

To gain a clearer understanding of the performance of various algorithms, \Cref{fig:xeon_pardegree-cyclesinstr} illustrates the cycles-per-instruction throughput, as reported by the \perf profiler, in relation to the degree of parallelism. A lower number of cycles per instruction indicates higher instruction throughput and, consequently, enhanced energy efficiency per unit of time for the tested algorithms. Among these, Zuckerli and the $k^2$-tree exhibit the fewest cycles per instruction. However, it is important to note that these two formats have higher overall energy requirements; their more compact nature necessitates more instructions, resulting in longer completion times and increased energy consumption. We believe that the lower cycles per instruction observed are due to the greater memory requirements of grammar-based solutions, which lead to more cache accesses, potentially causing pipeline stalls and increased latencies at the hardware level.

Interestingly, the throughputs of Zuckerli and, in particular, the $k^2$-tree tend to decline as the number of threads increases. In contrast, for the grammar-based approaches, the number of cycles required for each instruction initially follows a monotonically decreasing trend up to a parallelism degree of 8, followed by an increase up to 24 threads, and then decreases again.

\begin{figure*}
    \captionsetup[subfigure]{labelformat=empty}
    \centering

    \begin{subfigure}[t]{0.32\textwidth}
        \caption{eu-2005}
        \includegraphics[width=\textwidth]{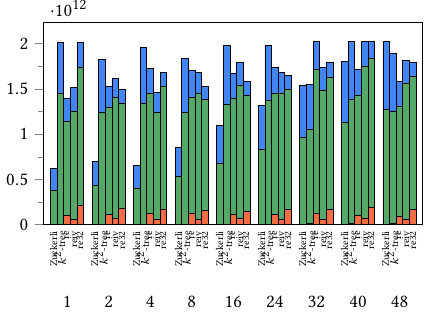}
    \end{subfigure}
    \hfill
    \begin{subfigure}[t]{0.32\textwidth}
        \caption{hollywood-2009}
        \includegraphics[width=\textwidth]{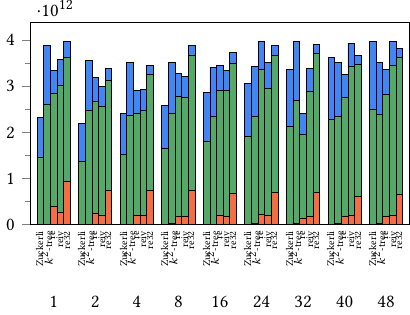}
    \end{subfigure}
    \hfill
    \begin{subfigure}[t]{0.32\textwidth}
        \caption{in-2004}
        \includegraphics[width=\textwidth]{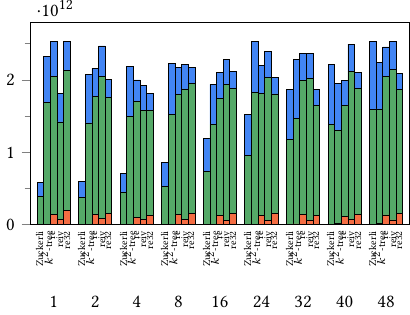}
    \end{subfigure}

    \vspace{1em} 

    \begin{subfigure}[t]{0.32\textwidth}
        \caption{ljournal-2008}
        \includegraphics[width=\textwidth]{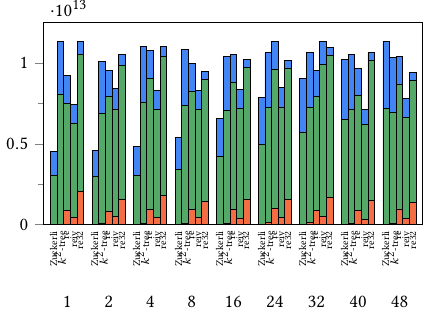}
    \end{subfigure}
    \hfill
    \begin{subfigure}[t]{0.32\textwidth}
        \caption{indochina-2004}
        \includegraphics[width=\textwidth]{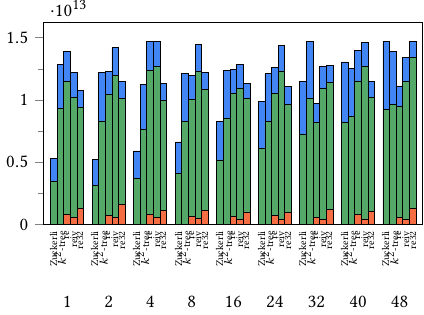}
    \end{subfigure}
    \hfill
    \begin{subfigure}[t]{0.32\textwidth}
        \caption{uk-2002}
        \includegraphics[width=\textwidth]{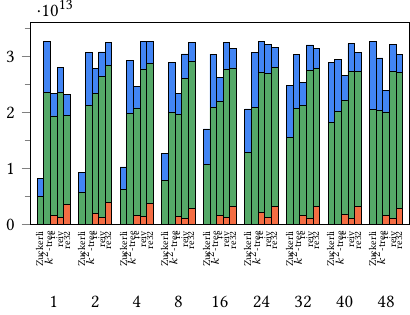}
    \end{subfigure}

    \vspace{1em} 

    \begin{subfigure}[t]{0.32\textwidth}
        \caption{arabic-2005}
        \includegraphics[width=\textwidth]{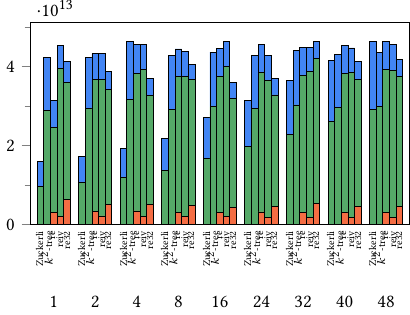}
    \end{subfigure}
    \hfill
    \begin{subfigure}[t]{0.32\textwidth}
        \caption{uk-2005}
        \includegraphics[width=\textwidth]{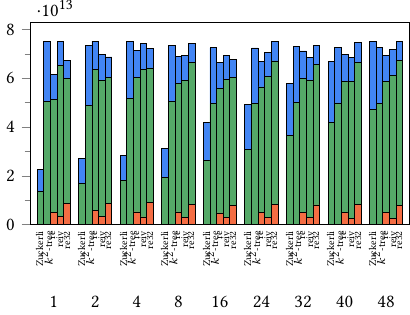}
    \end{subfigure}
    \hfill
    \begin{subfigure}[t]{0.32\textwidth}
        \caption{it-2004}
        \includegraphics[width=\textwidth]{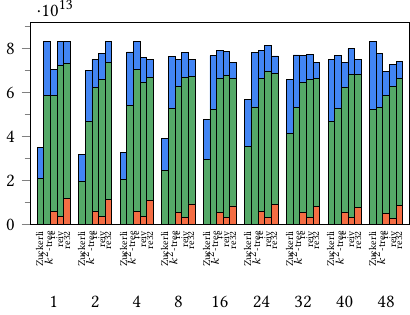}
    \end{subfigure}

\vspace{0.6cm}

\caption{Accesses to L1d cache memories on the \xeon Gold (y-axis) as a function of the parallelism degree (x-axis).}\label{fig:L1D}
\end{figure*}

\Cref{fig:L1D} illustrates the number of accesses to L1 data caches (L1D) for each matrix format as a function of the degree of parallelism. The total height of each bar represents the overall number of operations in the L1D cache, calculated by summing the number of load cache misses (red), load cache hits (green), and store operations (blue). The x-axis denotes the degree of parallelism, with each group of five bars corresponding to the five tested compression algorithms, presented in the following left-to-right order: Zuckerli, $k^2$-tree, \reans{}, \reiv{}, and \reint{}.

More succinct formats, such as Zuckerli, require fewer cache loads, resulting in fewer cache misses. The number of cache operations for Zuckerli increases with the degree of parallelism. In contrast, \reiv{} and \reint{} often exhibit non-monotonic behaviors; for example, in the \hollywoodX dataset, the number of operations for \reint{} decreases when scaling from 1 to 2 threads, but then increases again at least up to 32 threads. Similarly, for the \hollywoodX dataset, the number of operations for \reiv{} and \reans{} decreases initially, reaching a minimum at 4 threads, after which they rise again. The number of operations for the $k^2$-tree also follows a non-monotonic pattern, decreasing up to 24 threads.

\begin{figure*}
    \captionsetup[subfigure]{labelformat=empty}
    \centering

    \begin{subfigure}[t]{0.32\textwidth}
        \caption{eu-2005}
        \includegraphics[width=\textwidth]{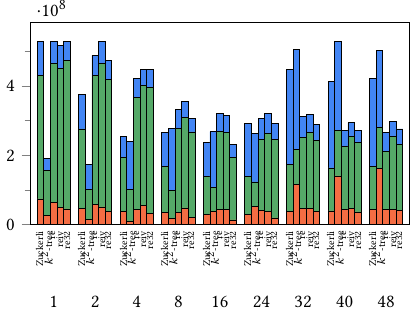}
    \end{subfigure}
    \hfill
    \begin{subfigure}[t]{0.32\textwidth}
        \caption{hollywood-2009}
        \includegraphics[width=\textwidth]{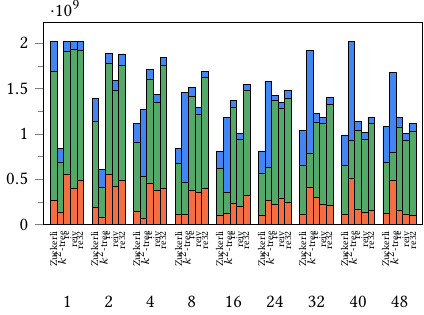}
    \end{subfigure}
    \hfill
    \begin{subfigure}[t]{0.32\textwidth}
        \caption{in-2004}
        \includegraphics[width=\textwidth]{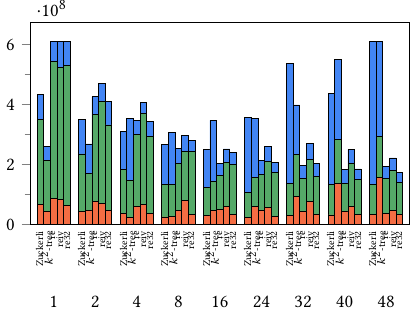}
    \end{subfigure}

    \vspace{1em} 

    \begin{subfigure}[t]{0.32\textwidth}
        \caption{ljournal-2008}
        \includegraphics[width=\textwidth]{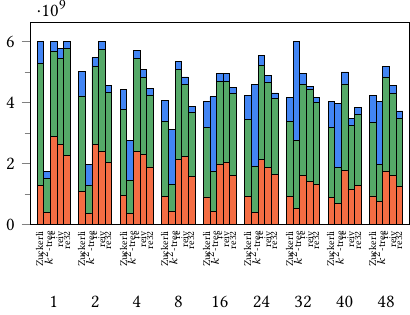}
    \end{subfigure}
    \hfill
    \begin{subfigure}[t]{0.32\textwidth}
        \caption{indochina-2004}
        \includegraphics[width=\textwidth]{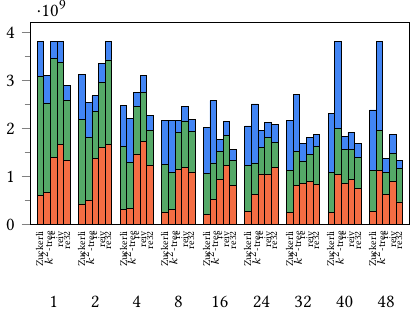}
    \end{subfigure}
    \hfill
    \begin{subfigure}[t]{0.32\textwidth}
        \caption{uk-2002}
        \includegraphics[width=\textwidth]{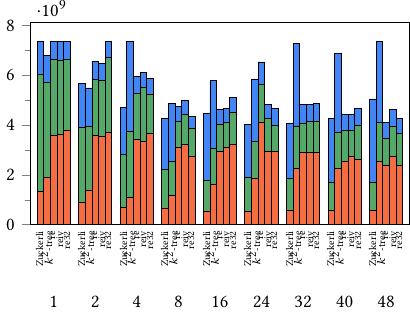}
    \end{subfigure}

    \vspace{1em} 

    \begin{subfigure}[t]{0.32\textwidth}
        \caption{arabic-2005}
        \includegraphics[width=\textwidth]{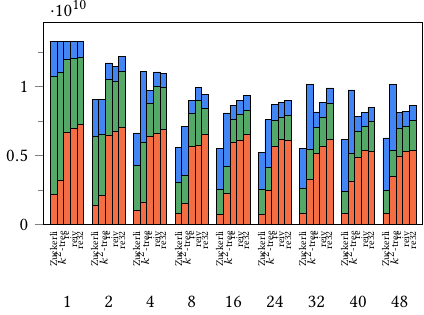}
    \end{subfigure}
    \hfill
    \begin{subfigure}[t]{0.32\textwidth}
        \caption{uk-2005}
        \includegraphics[width=\textwidth]{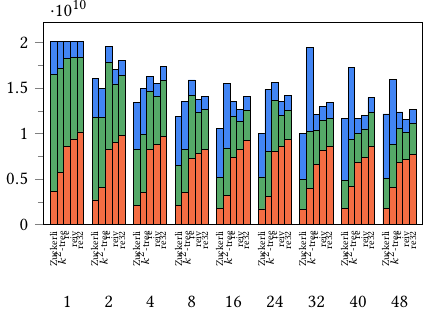}
    \end{subfigure}
    \hfill
    \begin{subfigure}[t]{0.32\textwidth}
        \caption{it-2004}
        \includegraphics[width=\textwidth]{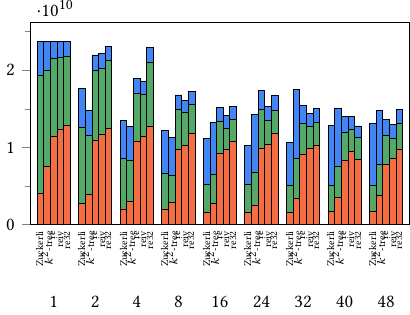}
    \end{subfigure}

\caption{Number of accesses to L3 cache memories on the \xeon Gold (y-axis) as a function of the parallelism degree (x-axis).}\label{fig:L3}
\end{figure*}

Similarly, \Cref{fig:L3} shows the number of accesses to the L3 caches. In this case, we observe for most datasets and formats a reduction in the number of accesses as the degree of parallelism increases. 

As noted in \cite{ladra-green}, the efficiency of cache accesses can help explain instruction throughput. Algorithms that make better use of cache hierarchies ---specifically those that predominantly reside in L1 caches--- exhibit higher instruction throughput compared to those that heavily rely on L3 caches. For instance, in the PageRank computations performed on the $k^2$-tree for the dataset \ljournalX, we observe that the number of L3 cache accesses increases monotonically up to 32 threads and then decreases, mirroring the behavior shown in \Cref{fig:xeon_pardegree-cyclesinstr}.
For the same dataset, when analyzing Zuckerli, we observe a decrease in L3 cache accesses, followed by an increase after 16 threads, a subsequent decline at 32 and 40 threads, and a final increase at a parallelism degree of 48. This pattern closely resembles the instruction throughput depicted in \Cref{fig:xeon_pardegree-cyclesinstr}.

This relationship between L3 accesses and cycles per instruction appears also in other datasets, particularly for Zuckerli, which is the slowest algorithm. Furthermore, we notice that the overall decrease in L3 accesses for Zuckerli often corresponds to an increase in L1D accesses, as shown in \Cref{fig:L1D}. This suggests that Zuckerli achieves higher instruction throughput by utilizing L1 caches more effectively than L3 caches.

  \begin{figure*}[t]
    \centering
    \includestandalone[width=\textwidth,mode=buildnew]{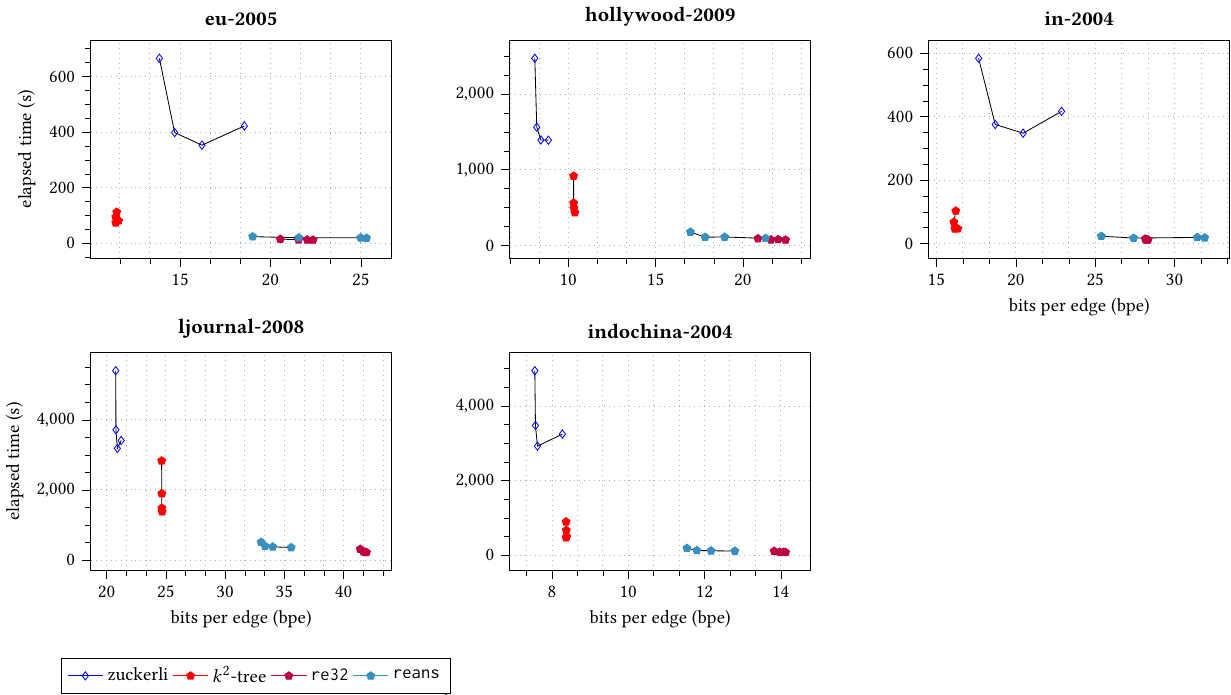}
    \caption{Bits per edge (x-axis) and elapsed times (y-axis) for 100 iterations of PageRank on the Raspberry Pi, for 1, 2, 4, and 8 threads.}
    \label{fig:raspi_bpe-elapsed}
  \end{figure*}

\subsection{Experiments on the Raspberry Pi}

To validate our results, we repeated the PageRank computations on an ARM-based Raspberry Pi (see the specifics in \Cref{subsec:archi}). Due to the limitations of this architecture, we only considered the five smallest graphs and ran the algorithms with only 1, 2, 4, and 8 threads. In addition, since we noticed in \Cref{fig:xeon_bpe-elapsed} that \reiv{} and \reans{} exhibit similar behaviors, we included only \reans{} in this set of experiments.

\Cref{fig:raspi_bpe-elapsed} illustrates the space--time performances on the Raspberry Pi. We observe that time and space are not always monotonic when scaling from 4 to 8 threads, as PMU and completion time increase, notably for Zuckerli. In three out of five datasets, Zuckerly is the most compact, while in the remaining two datasets, the $k^2$-tree demonstrates superior compactness. Regarding time performance, the ranking positions the grammar-based solutions \reint{} and \reans{} as the fastest, followed by the $k^2$-tree, with Zuckerly trailing behind. The grammar-based solutions achieve speeds at least $2\times$ faster than the $k^2$-tree, albeit at the cost of significantly increased memory usage.

In comparing \Cref{fig:xeon_bpe-elapsed} with \Cref{fig:raspi_bpe-elapsed}, we observe that on the Raspberry PI, even the single-threaded version of the $k^2$-tree has faster completion time than all Zuckerli parameterizations across all datasets: on the \xeon for \hollywoodX and \ljournalX Zuckerli with 4 threads resulted in faster computations than the single thread $k^2$-tree. 

\begin{figure*}
    \captionsetup[subfigure]{labelformat=empty}
    \centering

    \begin{subfigure}[t]{0.32\textwidth}
        \caption{eu-2005}
        \includegraphics[width=\textwidth]{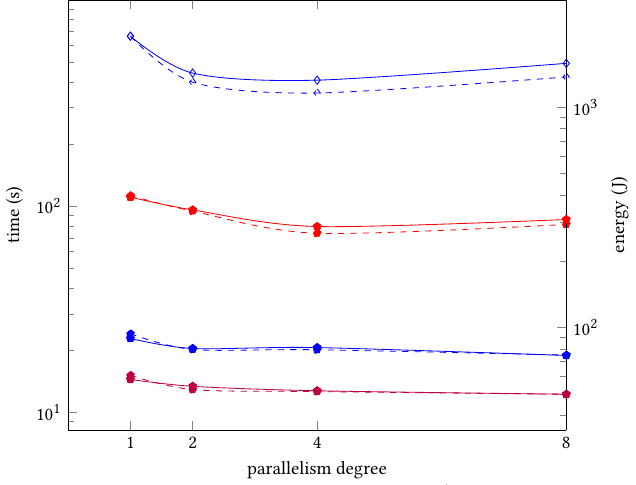}
    \end{subfigure}
    \hfill
    \begin{subfigure}[t]{0.32\textwidth}
        \caption{hollywood-2009}
        \includegraphics[width=\textwidth]{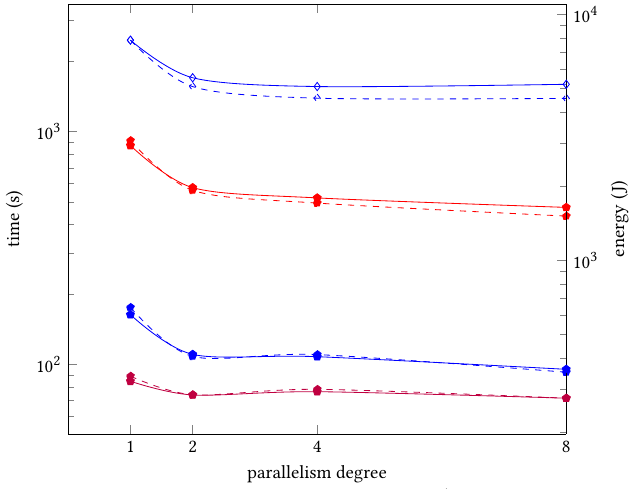}
    \end{subfigure}
    \hfill
    \begin{subfigure}[t]{0.32\textwidth}
        \caption{in-2004}
        \includegraphics[width=\textwidth]{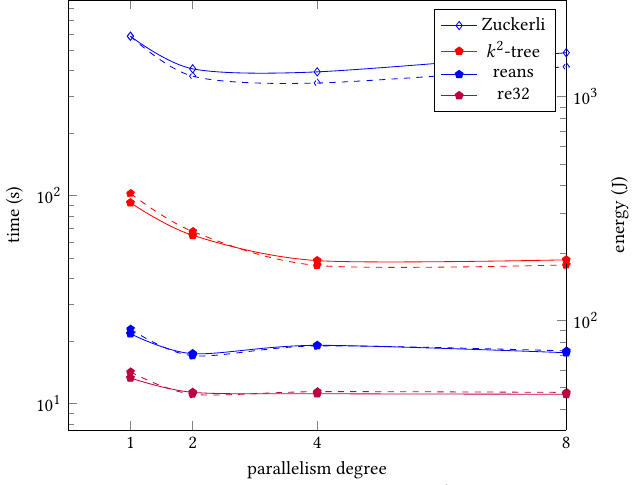}
    \end{subfigure}

    \vspace{1em} 

    \begin{subfigure}[t]{0.32\textwidth}
        \caption{ljournal-2008}
        \includegraphics[width=\textwidth]{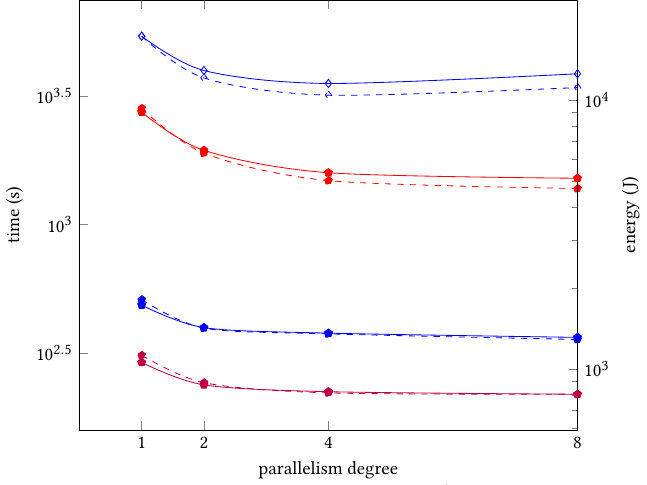}
    \end{subfigure}
    \begin{subfigure}[t]{0.32\textwidth}
        \caption{indochina-2004}
        \includegraphics[width=\textwidth]{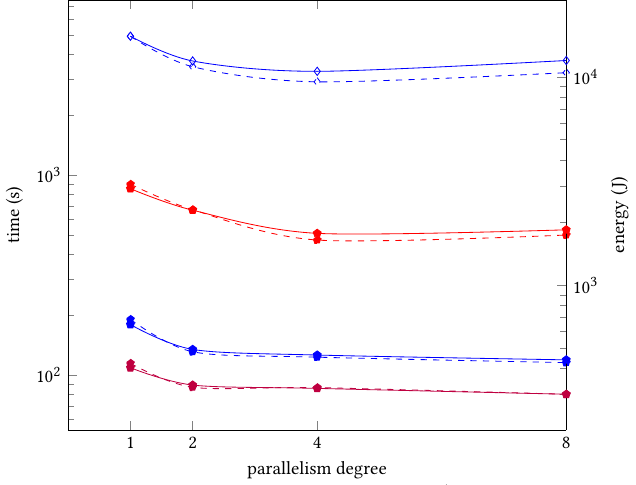}
    \end{subfigure}
    \hfill 

\vspace{0.6cm}

\caption{Time performance in seconds (dashed line; left y-axis) and energy requirements in Joules (solid line; right y-axis) on the Raspberry Pi 4 as a function of the number of threads (x-axis).}\label{fig:optimality_raspi}
\end{figure*}

\Cref{fig:optimality_raspi} shows time and energy usage as a function of the number of threads. As we mentioned in \Cref{subsec:power} since RAPL data is not available on the Raspberry Pi, the energy consumption was estimated by measuring the input current during the computation. This substantially different measurement technique yields results that align well with those for the \xeon, considering that we used at most 8 threads since the Raspberry has only 4 cores. Since the time-energy behavior in \Cref{fig:optimality_raspi} substantially confirms those of \Cref{fig:optimality_xeon}, we omit reporting the metrics we captured for L1d caches for the Raspberry Pi (keep in mind that the Raspberry Pi does not have any L3 caches).

Even for the Raspberry Pi, we observe a convex trend, particularly with Zuckerli. In this case, allocating additional threads (8 threads, exceeding the 4 available cores) prolongs completion time and increases resource utilization, resulting in greater energy inefficiency. In contrast, the data for Zuckerli on the \xeon machine in \Cref{fig:optimality_xeon} exhibit a less pronounced convex trend when scaling to high levels of parallelism. We believe that, in the server environment, when approaching the optimal parallelism degree, the increased energy required to operate with more threads offsets the benefit of slightly faster execution gained from the time speedup.
Overall our results suggest that the qualitative behavior observed on the \xeon server can be extended to resource-constrained edge devices.

\section{Conclusion and Future Work}\label{sec:future}

We have tested the running time, disk occupancy, memory usage, and energy consumption of three different algorithms for binary SpMVs, using them for computing the PageRank of large Web graphs and social networks. The main lessons we learned from our study can be summarized as follows. 

\begin{itemize}
    \item By employing an appropriate compressed representation, we can effectively tackle problems involving large datasets, even on resource-constrained devices.
    
    \item Different algorithms exhibit different space-time tradeoffs, which can be significantly influenced by the number of available threads.
    
    \item For PageRank computations, the $k^2$-tree appears to be the safe middle-ground choice: it is nearly as fast as grammar-based compressors and almost as space-efficient as Zukerli. Additionally, it shows the least deterioration in compression ratio as the number of threads increases.
    
    \item The choice of compressed representation should take into account the type of input data. All our representations were considerably less effective on the two social network graphs. However, the results for {\tt gzip} and {\tt xz} indicate that even those matrices are significantly compressible, suggesting that other compressed representations, such as those mentioned in \Cref{sec:other}, should be considered.
    
    \item Energy consumption is not solely proportional to running time, therefore it should be measured independently in critical scenarios. This is especially evident when considering the number of threads: our experiments show that the optimal number of threads for minimizing energy usage may differ from that for optimizing running time. We believe that further research in this area could lead to new theoretical challenges regarding the optimization of energy and the implementation of multicriteria time-energy solutions.
    
    \item The experiments demonstrate that by carefully selecting the compressed representation, energy usage can be reduced by one or two orders of magnitude. This reduction is observed in both server scenarios and when using a single-board Raspberry Pi.
    
    \item Regarding question Q3, our results presented in \Cref{fig:L1D} and \Cref{fig:L3} indicate that an increase in L1 and L3 cache operations impacts the cycles per instruction rate shown in \Cref{fig:xeon_pardegree-cyclesinstr}. This, in turn, affects overall time and energy performance, providing evidence that inadequate utilization of the cache hierarchy can degrade instruction-per-clock-cycle throughput, leading to increased latencies and energy inefficiencies.
    
\end{itemize}

\noindent For future work, we plan to explore additional lossless compression formats for matrices and vectors and investigate applications beyond PageRank. We believe that in-depth research in this area can uncover new theoretical challenges related to optimizing the energy-time tradeoff, providing valuable insights for software engineers seeking to reduce carbon footprints and extend the battery life of their solutions.

We also intend to study cross-platform energy-efficient implementations of other major compressed data structures, such as the FM-index, rank and select support structures, suffix arrays, and succinct tree topologies. This work will demonstrate the potential of data compression to reduce carbon footprints across diverse platforms and applications.

In the context of ML optimization, we propose exploring combinations of lossless tools with state-of-the-art lossy compression strategies. This approach aims to strike a balance between space and accuracy, allowing us to leverage the strengths of both methods and evaluate their effectiveness.


\bibliographystyle{IEEEtran}
\bibliography{access.bib}

\EOD

\end{document}